\PassOptionsToPackage{table,xcdraw}{xcolor}
\documentclass[format=acmsmall, review=false, anonymous = false, screen=true]{acmart}
\usepackage{booktabs} 
\usepackage{multirow}
\usepackage{graphicx}
\usepackage{threeparttable}
\usepackage{tabularx}
\usepackage{subcaption}
\usepackage{array}
\usepackage[skip=0.5\baselineskip]{caption}
\usepackage{hyperref} 

\setcopyright{acmlicensed}
\acmJournal{TOCHI}
\acmYear{2020} \acmVolume{1} \acmNumber{1} \acmArticle{1} \acmMonth{1} \acmPrice{15.00}\acmDOI{10.1145/3381804}

\usepackage[utf8]{inputenc}
\usepackage[english]{babel}

\begin{document}
\title[Using an AI-Powered Chatbot to Conduct Conversational Surveys]{Tell Me About Yourself: Using an AI-Powered Chatbot to Conduct Conversational Surveys with Open-ended Questions}

\author{Ziang Xiao}
\affiliation{%
  \institution{University of Illinois at Urbana-Champaign}
  \city{Urbana}
  \state{IL}
  \country{USA}
}
\email{zxiao5@illinois.edu}

\author{Michelle X. Zhou}
\affiliation{%
  \institution{Juji. Inc.}
  \city{San Jose}
  \state{CA}
  \country{USA}
}
\email{mzhou@acm.org}

\author{Q. Vera Liao}
\affiliation{%
  \institution{IBM Research AI}
  \city{Yorktown Heights}
  \state{NY}
  \country{USA}
}
\email{vera.liao@ibm.com}

\author{Gloria Mark}
\affiliation{%
  \institution{University of California, Irvine}
  \city{Irvine}
  \state{CA}
  \country{USA}
}
\email{gmark@uci.edu}

\author{Changyan Chi}
\affiliation{%
  \institution{Juji. Inc.}
  \city{San Jose}
  \state{CA}
  \country{USA}
}
\email{tchi@juji-inc.com}

\author{Wenxi Chen}
\affiliation{%
  \institution{Juji. Inc.}
  \city{San Jose}
  \state{CA}
  \country{USA}
}
\email{wchen@juji-inc.com}
\author{Huahai Yang}
\affiliation{%
  \institution{Juji. Inc.}
  \city{San Jose}
  \state{CA}
  \country{USA}
}
\email{hyang@juji-inc.com}

\renewcommand{\shortauthors}{Xiao et al.}

\begin{abstract}
The rise of increasingly more powerful chatbots offers a new way to
collect information through conversational surveys, where a chatbot
asks open-ended questions, interprets a user's free-text responses,
and probes answers whenever needed. To investigate the effectiveness and
limitations of such a chatbot in conducting surveys, we conducted a
field study involving about 600 participants. In this study with mostly open-ended
questions, half of the participants took a typical online survey on Qualtrics and the
other half interacted with an AI-powered chatbot to complete a
conversational survey. Our detailed analysis of over 5200 free-text responses
revealed that the chatbot drove a significantly higher level of
participant engagement and elicited significantly better quality
responses measured by Gricean Maxims in terms of their informativeness, relevance, specificity, and
clarity. Based on our results, we discuss design implications for
creating AI-powered chatbots to conduct effective surveys and beyond.
\end{abstract}



\settopmatter{printfolios=true}
\maketitle

\section{Introduction}
\label{sec:Intro}
\begin{figure}[t]
\centering
\includegraphics[width = 6cm]{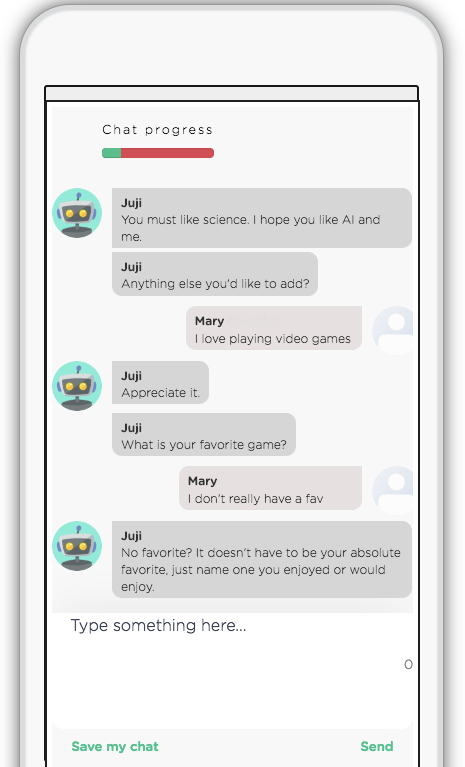}
\caption{A screenshot of a chatbot survey in our study.}
\label{fig:openning}
\end{figure}

In many domains, including HCI research \cite{muller2014survey},
conducting surveys is a key method to collect data. With the widespread use
of the internet, self-administered online surveys have replaced 
old-fashioned paper-and-pencil surveys and have become one of the most
widely used methods to collect information from a target audience
\cite{evans2005value,fricker2002advantages}. Compared to
paper-and-pencil surveys, online surveys offer several distinct
advantages. First, an online survey is available 24x7 for a target
audience to access and complete at their own pace. Second, it can
reach a wide audience online regardless of their geographic
locations. Third, online survey tools automatically tally survey
results, which minimizes the effort and errors in processing the results.

Due to the extensive use of online surveys, survey fatigue is now a
challenge faced by anyone who wishes to collect data. Research indicates two typical types of survey
fatigue \cite{porter2004overcoming}. One is \textit{survey response
fatigue}. Since people are inundated with survey requests, they are
unwilling to take any surveys \cite{porter2004raising}. The other is \textit{survey-taking
fatigue}. Evidence shows that as a survey grows in length,
participants spend less time on each question and the completion rate
also drops significantly \cite{ben2008respondent,porter2004multiple}. For example, one of
the biggest survey platforms, SurveyMonkey, shows that on average,
participants spend 5 minutes to complete a 10-question survey but 10
minutes to finish a 30-question survey \footnote{https://www.surveymonkey.com/curiosity/5-best-ways-to-get-survey-data/}.

The problem is exacerbated with open-ended questions because of the extra time and effort required for formulating and typing responses to such questions \cite{ben2008respondent,
chen2017effects}. Open-ended questions are an important method to collect
valuable data and are widely used in self-administered online
surveys \cite{muller2014survey}. In particular, open-ended questions
allow respondents to phrase their answers freely when the options of responses cannot be pre-defined or the pre-defined responses may introduce biases  \cite{lavrakas2008encyclopedia, chen2017effects}. Moreover,
open-ended questions help collect deeper insights, such as the background and rationales behind the
answers \cite{singer2017some, chen2017effects}. However, open-ended questions
often induce more cognitive burdens and survey-taking fatigue, and
respondents are more likely to skip such questions or provide
low-quality or even irrelevant answers \cite{ben2008respondent,
reja2003open, chen2017effects}. Consequently, survey-taking fatigue adversely
affects the quality and reliability of the data collected especially
when open-ended questions are involved \cite{conrad2005interactive,
oudejans2010using, chen2017effects}

To combat survey taking fatigue especially to motivate and guide
survey participants to provide quality answers to open-ended
questions, several approaches have been proposed. One set of proposals
is to inject interactive features into an online static survey, such
as providing response feedback \cite{conrad2005interactive} and
probing responses \cite{oudejans2010using}, to improve response
quality and encourage participant engagement. However, no existing
survey platforms support such interactive features nor do they
automatically motivate and guide survey participants to provide
quality answers to open-ended questions during a survey.

A lack of support of such interaction features on existing platforms
may be due to two main reasons. First, it is difficult to
automatically interpret participants' natural language responses to
an open-ended question due to the diversity and
complexity of such responses \cite{chen2017effects}. For example, when asked \textit{``What do
you think of the product''}, participants' responses could be:  \textit{``N/A''},  \textit{``I don't know''}, or  \textit{``Although I've
heard of the product, I've never used it so I don't know what to
say.''} Interpreting such highly diverse or complex free-text input
requires sophisticated natural language processing algorithms, which is a
non-trivial task \cite{grudin2019chatbots}. Second, even if a system can
interpret participants' free-text responses to open-ended questions,
it is difficult to manage potentially complex interactions
based on participant responses. Using the above example, a
participant may be unwilling to answer the open-ended question and
may even provide a gibberish response such as  \textit{``afasf asfasf afyiasfaf
asf''} \cite{gadiraju2015understanding}. In another example, a participant is
willing to answer the question, but provides a very terse
answer such as  \textit{``not bad''} as opposed to detailed, rich
information. Yet in another example, instead of answering the
question, a participant asks a clarification question  \textit{``Which
aspects of the product do you want me to comment on''}. Handling all
these situations or their compositions requires that a system not only
understands a participant's input but also automatically handles diverse interaction situations, which is very challenging to
implement \cite{grudin2019chatbots}.

On the other hand, the advent of chatbots with their increasingly more
powerful conversational capabilities can offer an alternative approach
to static online surveys. Specifically, an AI-powered chatbot can
conduct a \emph{conversational survey}. As shown in Fig \ref{fig:openning}, in a conversational survey, a chatbot asks
open-ended questions, probes answers, and handles social dialogues.

Intuitively, a chatbot-powered conversational survey
retains the advantages of online surveys and offers several additional
benefits especially facilitating gathering participant responses to
open-ended questions. First,
a chatbot can frame survey questions in more personalized,
conversational messages, which might help improve participant
engagement and response quality \cite {chen2017effects, heerwegh2006personalizing, krosnick1999survey}. 
Second, the perceived anthropomorphic characteristics of a chatbot can potentially deliver
human-like social interactions that encourage survey participants to
reveal personal insights \cite{tallyn2018ethnobot}. Third, it is
natural for a chatbot to interactively encourage information exchange
in the course of a survey, such as providing response feedback and
probing responses, which in turn helps reduce survey-taking fatigue
and improve response quality. Moreover, it is the inherent
functions of chatbots that interpret diverse user natural language input and handle
complex conversations. As a result, chatbots can potentially
serve as a moderator and proactively manage a survey process, such as
dealing with ``uncooperative'' participants, clarifying the meaning of
a question per a participant's request, and guiding a participant to
provide richer and more authentic responses \cite{li2017confiding, sundar2019machine}.

Despite their benefits, chatbots bear several risks for their use in
conducting surveys. First, a turn-by-turn chat requires participants
to take extra time and effort to complete a survey. It is unclear
whether people would be willing to take the time to chat and complete
a survey, let alone providing quality responses. The risk is even
higher for surveys with paid participants, who would not be rewarded
for taking a longer survey. Second, current chatbots are far from
perfect and their limited conversation capabilities may lead to user
disappointment and frustration \cite{grudin2019chatbots}. It is unknown
whether the limited capabilities would deter participants from
offering quality responses or completing a survey. Moreover, it is
difficult for a chatbot to accurately interpret and properly respond
to humans' diverse free-text input to open-ended questions
\cite{devault2014simsensei}. Once participants realize that a chatbot
cannot fully understand or assess their responses, it is unknown
whether they would do mischief by intentionally feeding the chatbot
with bogus responses, which would adversely affect the overall
response quality. Finally, the use of a personified conversational
system may lead to user behaviors that affect survey quality. For
example, studies show that people have positivity bias when giving
opinions to an agent \cite{thomaz2008teachable}, producing
potentially biased survey results.

To our knowledge, there have not been any in-depth studies examining
the effectiveness and limitations of AI-powered chatbot surveys in
contrast to typical online surveys. We, therefore, ask two research
questions:
\begin{itemize}
        \item {\textbf{RQ1:} How would user response quality differ,
        especially the quality of user free-text responses to
        open-ended questions in an AI-powered chatbot-driven
        survey vs. a traditional online survey?}  \vspace{0.5mm}
        
        \item{\textbf{RQ2:} How would participant engagement
        differ in an AI-powered chatbot-driven survey vs. a
        traditional online survey?}
\end{itemize}

To answer the above research questions, we designed
and conducted a field study that compared the use of an AI-powered
chatbot versus a typical online survey with the focus on eliciting
user answers to open-ended questions. As mentioned above, there are potential
benefits and risks of using chatbots to conduct surveys, especially
when involving open-ended questions. However, none of the benefits or
risks have been examined. In this first study, we thus decided to
focus on examining the \emph{holistic} effect of a chatbot instead of
investigating the effect of separate chatbot features.

Additionally, to ensure that our study is based on real-world survey
practices and offers practical value, we collaborated with a
global-leading market research firm that specializes in discovering
customer insights for the game and entertainment industry. Per the
firm's request, our field study was to learn how gamers think and feel
about two newly released game trailers. The study involved about 600
gamers, half of whom took a chatbot survey and the other half filled
out a typical online survey. Through detailed analyses of over 5000
collected responses, we addressed our two research questions. We
also discussed the design implications for creating effective chatbots
to conduct engaging surveys and beyond.

To the best of our knowledge, our work is the first that
systematically compared the holistic effect of an
AI-powered conversational survey with that of a typical online survey
on response quality and participant engagement. As a result, our work
reported here provides three unique contributions.

\begin{enumerate}

\item {\textit{ \textbf{An understanding of the holistic effect of
AI-powered chatbots on users in surveys with open-ended questions.}} Our findings
revealed the practical value of AI-powered chatbot surveys
especially in eliciting higher quality responses and increasing
respondents' engagement.}

\vspace{1mm}

\item {\textit{ \textbf {Design implications of AI-powered chatbots for
survey success.}} Our work discusses design considerations, such as
enabling active listening and early intervention, for creating
effective chatbots for conducting conversational surveys, especially
facilitating the collection of quality survey responses and improving
participant engagement.}
\vspace{1mm}

\item {\textit{\textbf{New opportunities of conducting AI-enabled, human-subject research}}. The demonstrated effectiveness of an AI-powered chatbot survey and the simplicity of creating such a
chatbot-driven survey open up opportunities of employing AI-powered chatbots
to aid in human-subject research, including AI-powered semi-structured
interviews and chatbot-driven longitudinal field studies.}
\end{enumerate}

\section{Related Work}
\label{sec:Relatedwork}
Broadly, our work is related to research in six areas as detailed below. 
\vspace{-2mm}
\subsection{Conversational AI and Chatbots}
\label{sec:ConversationalAIandChatbots}

Our work is related to a rich body of studies on the applications of
conversational AI or chatbots. For example, past studies have examined
chatbots as a personal assistant~\cite{liao2018all}, intelligent
tutor~\cite{graesser2005autotutor}, customer service
agent~\cite{allen2001toward,xu2017new}, job
interviewer~\cite{li2017confiding}, and worker's
companion~\cite{williams2018supporting}.

The HCI community has long been interested in the interaction benefits
offered by conversational interfaces. The general consensus is that
conversational interfaces offer several advantages over traditional
WIMP (Windows, Icons, Menus, and Pointers) interfaces
\cite{brennan1990conversation, luger2016like}. First, conversational
interfaces offer a natural and familiar way for users to express
themselves, which in turn improves the usability of a system. Second,
such interfaces are flexible and can accommodate diverse user requests
without requiring users to follow a fixed
path \cite{traum2017computational}. Third, such interfaces are often
personified and their anthropomorphic features could help attract user
attention and gain user trust \cite{walkerhci1996}.

Inspired by the potential advantages of conversational interfaces
over WIMP-based user interfaces, our work investigates the use of
conversational interfaces for conducting surveys. Differing from
existing works on conversational interfaces, we are exploring a new
application of conversational AI for conducting surveys, which has its
own set of unique challenges. For example, survey participants would
not be as motivated or cooperative as job candidates who interact with
a conversational AI agent \cite{li2017confiding,
devault2014simsensei}. Neither would survey participants be as
tolerant or patient as a student or an employee who interacts with an
AI companion \cite{Xiao2019, williams2018supporting}.

Furthermore, conversational interfaces are far from perfect due to
technical difficulties in processing user natural language
expressions and managing diverse and complex conversation
situations \cite{grudin2019chatbots, traum2017computational}. It is thus unknown how the
shortcomings of conversational interfaces (e.g., failure to understand
a user's input during a survey) might influence survey participants
and survey results. Therefore, we hope to investigate whether and how
conversational interfaces might bring in practical values to
traditional survey practices, which use WIMP-based interfaces to
elicit information. As the first step, we compare the holistic effect
of a chatbot survey with that of a traditional online survey on survey
participants and survey results in a real-world setting.

\subsection {Conversational AI for Information Elicitation}
\label{sec:ConversationalAIforInformationElicitation}

Our work is directly relevant to the use of conversational AI for
information elicitation. Researchers have built various AI agents to
elicit information from a user through a one-on-one, text-based
conversation, such as eliciting information from a job candidate
\cite{li2017confiding} and gathering data from a study participant
\cite{tallyn2018ethnobot}. Information elicitation may serve various
purposes. A common task is to elicit ``parameters'' for information
retrieval or recommendation \cite{ricci2015recommender,
radlinski2017theoretical,trippas2018informing}. This kind of system
often supports task-oriented conversations, such as helping a customer
book a flight reservation, finding a desired restaurant, and
scheduling an event \cite{hemphill1990atis,mcglashan1992dialogue,
cranshaw2017calendar}. The main goal of such systems is for an agent
to gather required information (e.g., travel dates) to perform a
task~\cite{bohus2009ravenclaw}.

More recently, conversational AI has been explored as a means to elicit
information for additional purposes beyond fulfilling a specific
task. For example, there have been agents that elicit information for
recommending products, movies, and jobs \cite{kang2017understanding,
zhang2018towards}, group decision support \cite{shamekhi2018face},
psychotherapy \cite{schroeder2018pocket,williams2018supporting}, and
voting \cite{folstad2017chatbots}.  An emerging area is using
conversational AI to conduct in-depth interviews
\cite{devault2014simsensei, li2017confiding} or longitudinal studies
in the real world \cite{tallyn2018ethnobot,
williams2018supporting}. For example, Li et al. \cite{li2017confiding} deployed agents to
interview a pool of real job candidates and compared the effect of
two agent personalities on the candidates' behavior. Tallyn et al. \cite{tallyn2018ethnobot}
developed a chatbot to gather ethnographic data from participants in
real-time. They showed that a simple chatbot with little language
understanding capabilities was effective in capturing data from the
participants. In a more recent study, Xiao et al. \cite{Xiao2019} used a chatbot to interview university students and gather their thoughts and feelings about teaming. 

Different from our investigation of using a chatbot as a general surveying tool, prior studies tended to focus on examining the
feasibility and effectiveness of a chatbot for a specific
information elicitation task. For example, Xiao et al. studied the use
of a chatbot for eliciting student preferences and opinions about
team building and investigated whether and how the elicited information predicted team performance \cite{Xiao2019}. Li et al. built a chatbot to
elicit information from job candidates and examined the candidates'
trusting behavior with the chatbot \cite{li2017confiding}. However, unlike our work, 
none of the existing studies compared the effectiveness of a chatbot in
information elicitation with that of a traditional approach. In particular, we want to quantitatively measure
the holistic effect of chatbots on user engagement and response
quality compared to that of a traditional online survey.

Although a typical online survey can elicit information through
various question prompts, including open-ended questions, such a
survey is not interactive or adaptive in nature. For example, in such
a process, a survey participant cannot ask a clarification question,
neither can the system probe a user response. On the other hand,
chatbots can naturally employ a diverse set of conversation prompts
to elicit information interactively, such as question prompts
~\cite{zhang2018towards}, follow-up probes \cite{tallyn2018ethnobot},
and topic-specific discussions
~\cite{ricci2015recommender,radlinski2017theoretical}. In addition,
conversation prompts can be context sensitive. For example, Williams
et al. employed both emotion-centric prompts \textit{``how do you
feel''} and task-centric prompts \textit{``what did you do''} to
elicit rich responses from users about work experience to improve
workplace productivity \cite{williams2018supporting}.

However, existing work has not examined how a
chatbot's often imperfect conversation capabilities would affect
information elicitation involving open-ended questions, including user
response quality and satisfaction. A recent study showed that
most chatbots can hardly understand user input and manage seemingly
simple tasks such as scheduling a meeting \cite{grudin2019chatbots}. This is
because users' natural language expressions are highly diverse and
challenging to interpret. Moreover, a seemingly simple conversation is
often still complex and nonlinear (i.e., going back and forth with a
user on a topic), which requires a chatbot to continously track and
proactively manage a conversation context \cite{grudin2019chatbots,
traum2017computational}. Our study is thus set out to explore both the advantages
and shortcomings of a chatbot in conducting surveys with open-ended
questions and to compare its holistic effect against that of a
traditional online survey to answer our two research questions.

\subsection {Conversational AI and Information Disclosure}
\label{sec:ConversationalAIforInformationDisclosure}
In the context of information elicitation, studies show that
conversational AI may enhance
user engagement and encourage self-disclosure. Prior work has
demonstrated that the exhibited social behaviors of agents are
effective in improving user engagements in various social settings by
a set of metrics, such as interaction duration, breadth, and depth of
self-disclosure (e.g., discussing personal subjects), and a positive
attitude toward the agent and interaction outcome
\cite{bickmore2011relational,shamekhi2018face, sundar2019machine}.

On the other hand, user's positive attitude toward AI agents has raised
concerns on user privacy protection and encouraged studies on
investigating the effect of chatbots on user trust and privacy in the
context of information elicitation. For example, a recent study showed
that users trusted a customer service chatbot more if they were well-informed
in the conversation where the information was
stored \cite{folstad2017chatbots}. Sannon et. al found that users were
less likely to share personal sensitive information (e.g., financial
or health stress) with a personified
chatbot \cite{sannon2018personification}. However, none of the existing studies compared survey participants' behaviors (e.g., self-disclosure and answer
quality) influenced by a chatbot versus in a traditional online
survey. We thus set out to gauge how conversational AI affects user
engagement and the quality of survey results, hoping to discover new
survey methods that may improve traditional online survey practices.

\subsection {Evaluating Conversational Interfaces}
\label{sec:evaluatingconversationalinterfaces}

With the advent of conversational interfaces, evaluating the
effectiveness of such interfaces is a continuously evolving research
topic. Traditionally, the evaluation criteria have been divided into
objective metrics on the system performance (e.g., user input
interpretation accuracy and user task completion rate) and subjective
metrics based on user feedback (e.g., user satisfaction)
\cite{mctear2016conversational, walker1997paradise}. Objective
metrics are directly computed from logs of the interaction and can be
based on task or domain coverage, error rate, number of interaction
issues, accuracy or other metrics compared to ``ground
truth'' \cite{mctear2016conversational,
dybkjaer2004evaluation,liu2016not}.  Subjective metrics are often
based on user opinions around certain aspects, such as
satisfaction, and intelligibility, (e.g.,
~\cite{hone2000towards}). There are also comprehensive methodologies
that consider both subjective user satisfaction and objective
performance metrics including task success, dialog efficiency (e.g.,
duration, total turns) and dialog quality (e.g., latency)
\cite{walker1997paradise}.

In addition to examining user satisfaction and usability of
conversational AI agents, HCI researchers have investigated how agent
behavior impacts users' social perceptions, such as
trust \cite{cassell2000external},
rapport \cite{bickmore2001relational, novick2014building},
anthropomorphism, and likability \cite{bartneck2009measurement}. Such
user subjective feedback is often measured by questionnaires, i.e.,
Likert-scale ratings on self-reported statements. Additionally,
automatic methods have been developed to predict user satisfaction based on
users' behavioral signals, such as dialogue acts, words, and user
actions \cite{jiang2015automatic,liao2018all}.

Compared to the existing work, our study focuses on evaluating the
outcomes of a conversational interface with a target goal--
collecting high-quality survey data and keeping the survey taker
engaged. We, therefore, have proposed several content-based metrics to
measure response quality and participant engagement.

\subsection {Conversational Interfaces vs. Graphical User Interfaces}
\label{sec:civsgui}

Our work is also related to evaluating the effect of a conversational
interface vs. that of a graphical user interface (GUI). A recent study
by YouGov compared the use of a traditional GUI form with a Facebook
Messenger Bot for conducting a survey \footnote{https://www.ama.org/publications/eNewsletters/Marketing-News-Weekly/Pages/why-chatbots-are-the-future-of-market-research.Aspx}. They found that
the chatbot significantly improved response rate. More recently,
researchers compared the response quality between a chatbot
survey and a form-based survey in more depth \cite{kimchi2019}. They
also found that the chatbot surveys produced more differentiated
responses and the participants were less likely to satisfice. However,
all the existing studies used only choice-based questions and have not
examined how chatbot-driven surveys would impact user responses to
open-ended questions, which has been one of the major challenges in
typical online surveys \cite{oudejans2010using}.

In other task domains, researchers have explored how a conversational
interface in lieu of a traditional GUI interface impact user
behavior. One such area is conversational search
\cite{trippas2018informing,thomas2018style}. For example, Trippas et
al. \cite{trippas2018informing} show that verbal communications led
to more complex user queries such as having multiple requests in one
utterance, while the interactivity encouraged user collaborative
behavior, such as actively requesting more specific information.

Similar to this line of work, we compare the effect of using a
conversational interface vs. a traditional GUI for conducting
surveys. However, we focus on quantifying their effect on the quality
of collected \textit{free-text} survey responses and user engagement,
which has never been addressed before.

\subsection {Improving Survey Quality}
\label{sec:surveyresearch}

Our work is related to survey research on improving survey
quality. Researchers have put tremendous effort into improving sample
validity and response quality. Heerwegh and Loosveldt
\cite{heerwegh2006personalizing} find that personalization can significantly increase web survey response rate by 4.4\% while not necessarily leading to social desirability response bias. Behr et al. \cite{behr2012} have tested three probing variants and found that such probings lead to a higher number of meaningful answers in web surveys. In a field experiment with over 4000 participants,
De Leeuw et al. \cite{deLeeuw2015} have shown that a polite probe can successfully reduce
item non-response (``don't know'') without sacrificing the reliability
of the final answers. Conrad et al. \cite{conrad2005interactive}
also show that interactive feedback can improve the quality of
responses in web surveys. Additionally,
Oudejans and Christian \cite{oudejans2010using} propose to use explanations and probings to
improve the quality of user responses to open-ended questions.

On the one hand, our work is built on top of existing findings. For
example, we learned that interactive features, such as response
feedback and probings, help improve participation and response
quality. On the other hand, our study is a natural extension of
existing efforts. In particular, we explore the use of chatbots to
offer various interactive features in a survey, hoping that such
features would improve participant engagement and response
quality.

\section{Study Method}
\label{sec:Method}

To answer our two research questions, we designed and conducted a
between-subjects field study that compared the outcomes of two survey
methods, an AI-powered chatbot survey and a typical form-based survey,
on the quality of collected information and participant engagement.

\subsection{Study Background}
\label{sec:Studybackground}

To ensure that our findings have ecological validity and practical value, we teamed up with a
global leading market research firm that specializes in discovering
customer insights for the entertainment industry, including game
companies and movie studios. Per the request of the firm, we set up
the field study to accomplish two goals. First, the firm wanted to
gauge gamers' opinions of two video game trailers recently released at
the Electronic Entertainment Expo (E3) 2018, the premier trade event
for the video game industry. Second, they wanted to compare the effect
of a chatbot survey with that of a typical online survey which they
frequently use to collect such information.

\subsection{Study Platform}

To compare the effect of a chatbot survey with that of a typical
form-based survey, our study was set up on two platforms.

\subsubsection{Qualtrics}

Qualtrics (\href{https://www.qualtrics.com/}{qualtrics.com}) is one
of the most popular online survey platforms. Since our collaborator
uses Qualtrics frequently for market research surveys, they set up the
form-based survey used in this study on Qualtrics. In the Qualtrics
survey, an open-ended question is presented with a text box where a
participant enters his/her answer. Participants can view and answer
only one question at a time and must submit their answers to a
presented question before moving on to the next question. A web URL was
generated to distribute the survey.

\begin{figure}[t]
  \begin{subfigure}[b]{0.90\textwidth}
    \includegraphics[width=\textwidth]{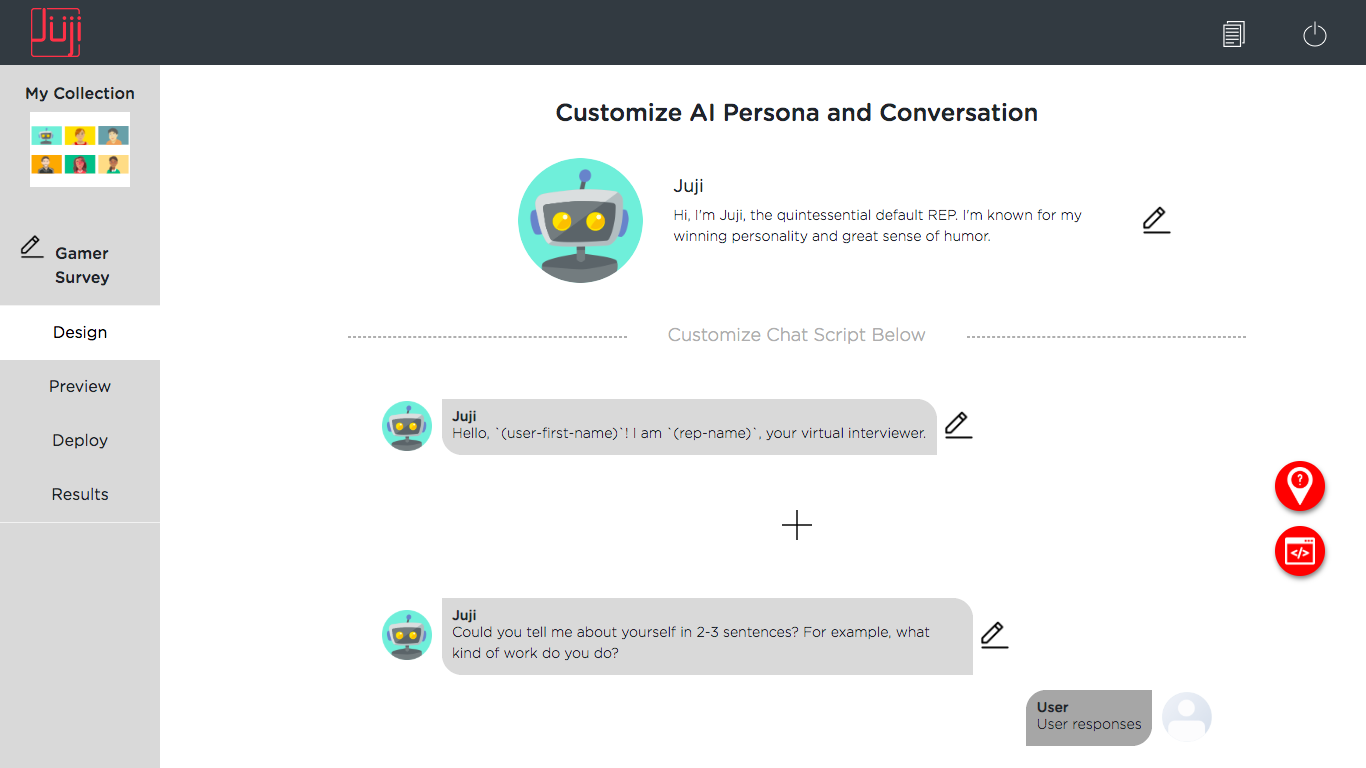}
    \caption{Juji's interface for editing questions}
    \label{fig:juji1}
  \end{subfigure}
  \begin{subfigure}[b]{0.90\textwidth}
    \includegraphics[width=\textwidth]{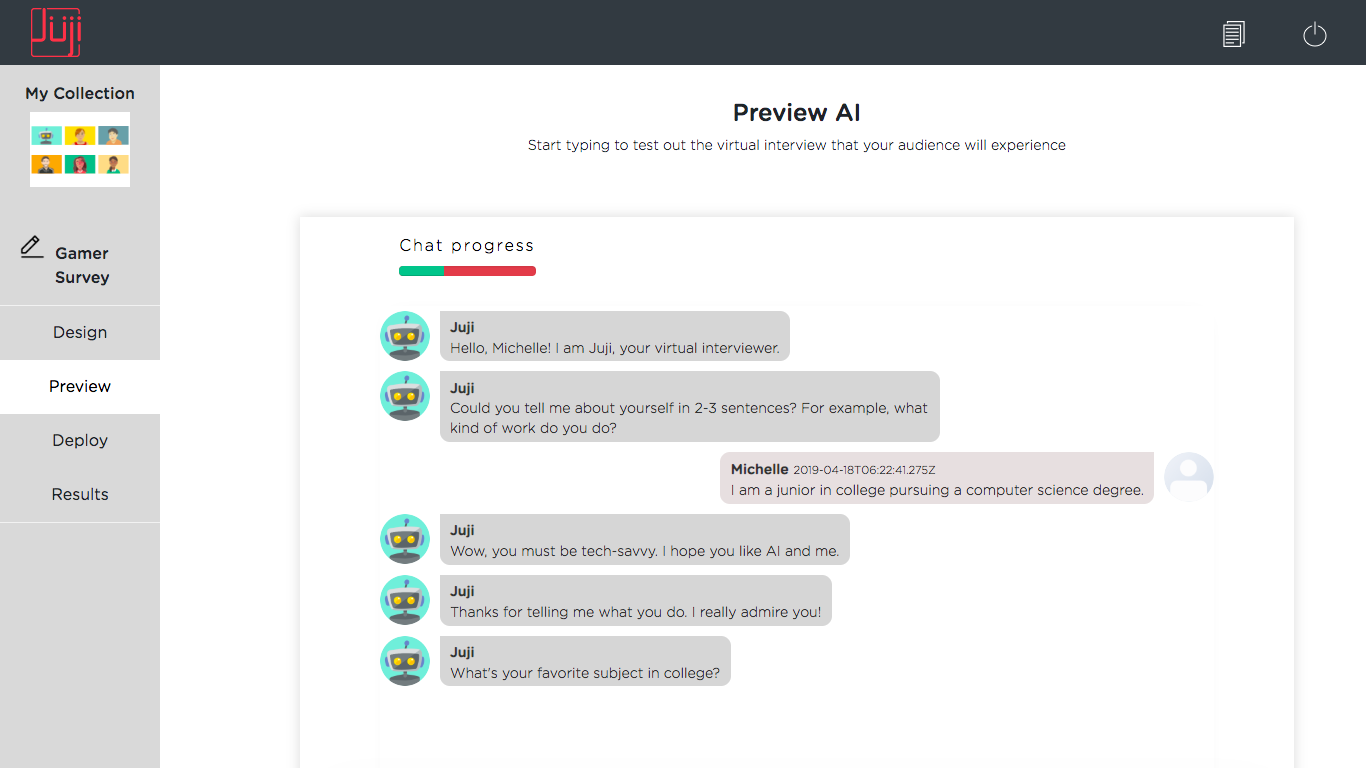}
    \caption{Juji's interface for previewing the chatbot}
    \label{fig:juji2}
  \end{subfigure}
  \caption{Juji's graphical user interface (GUI) for creating a chatbot survey.}
  \label{fig:Juji's GUI}
\end{figure}

\subsubsection{Juji Chatbot}

\begin{table}[t]
\centering
\begin{tabular}{m{20mm}p{50mm}m{60mm}}
\hline
\rowcolor[HTML]{EFEFEF}
\textbf{Technique} & \textbf{Synopsis}  & \textbf{Example}  \\ \hline
&&\\
\multirow{2}{20mm}{\textbf{Response Feedback}} & \multirow{2}{50mm}{Acknowledge a user's responses and provide positive feedback to encourage information disclosure} &  User: \textit{``I am a student majoring in Computer Science.''}                                 \\
   &  & Juji: \textit{``Wow, you must love science. I hope you like AI and me.''}  \\
 &&\\
\rowcolor[HTML]{EFEFEF} 
 &&\\
 \rowcolor[HTML]{EFEFEF} 
\textbf{Response Prompting} & Ask for more information       & Juji:\textit{``I'd love to hear more. Go on...''}   \\
\rowcolor[HTML]{EFEFEF} 
 &&\\
 &&\\
 &    & User: \textit{``I love to read books and cook.''}                                                         \\
\multirow{-2}{20mm}{\textbf{Response Probing}}  & \multirow{-2}{50mm}{Probe answers to elicit deeper information}                                                       & Juji: \textit{``Why do you like to read?''} \\
 &&\\
\rowcolor[HTML]{EFEFEF}  
 &&\\
 \rowcolor[HTML]{EFEFEF} 
 &    & User: \textit{``You are so smart''}   \\
\rowcolor[HTML]{EFEFEF}  
\multirow{-2.5}{22mm}{\textbf{Social Acknowledgement}}  & \multirow{-2.5}{45mm}{Acknowledge a user's input to convey understanding and verbalize emotions.}                                                       & Juji: \textit{``Thanks, I'm glad you are happy with me''} \\
\rowcolor[HTML]{EFEFEF}  
 &&\\
  &&\\
\multirow{2}{25mm}{\textbf{Handle Conversation Flow}} & \multirow{2}{45mm}{Handle a user's request for controlling an interview flow and encourage continuation.} &  User: \textit{``Wait. Give me a second''} \\
  &  & Juji: \textit{``Take your time. We'll continue whenever you are ready''}  \\
  &&\\
\rowcolor[HTML]{EFEFEF} 
  &&\\
 \rowcolor[HTML]{EFEFEF} 
 &    & User: \textit{``I don't know.''}   \\
\rowcolor[HTML]{EFEFEF}  
\multirow{-2.5}{20mm}{\textbf{Handle User Excuses}}  & \multirow{-2.5}{50mm}{Handle a user's excuses not to answer a question and encourage continuation.}                                                       & Juji: \textit{``No worries, just share what's on your mind.''} \\
\rowcolor[HTML]{EFEFEF} 
&&\\
&&\\
\multirow{2}{20mm}{\textbf{Handle User Questions}} & \multirow{2}{50mm}{Handle a user's question to encourage continuation.} &  User: \textit{``Why are you repeating the question?''} \\
  &  & Juji: \textit{``Why not? :) I am pretty sure this is for your own good since I am a responsible A.I.''}  \\
  &&\\
  \hline

\\
\end{tabular}
\caption{Juji chatbot's key conversational skills.}
\label{tab:conversationalfeatures}
\end{table}

Juji (\href{https://juji.io/}{juji.io}) offers an AI-powered chatbot
platform where users can create and deploy their own customized
chatbots for various tasks. For example, Juji was used to develop
chatbots for interviewing job candidates \cite{li2017confiding} or
interviewing college students for teaming purposes
\cite{Xiao2019}.

Specifically, a survey creator uses Juji's graphical user interface
(GUI) to input a set of survey questions and the order of the
questions to be asked. A chatbot is then automatically built with a
set of default conversation capabilities as described below, such as
handling a conversation around an open-ended question as well as
managing several types of user digressions or side-talking
dialogues. Fig \ref{fig:Juji's GUI} are screenshots of the Juji
creator GUI with which a survey creator can design, preview, and
deploy a chatbot to conduct a conversational survey. The survey
creator can add, delete, and modify a survey question (Fig \ref{fig:juji1}). Juji supports several types of questions, including
choice-based questions and open-ended questions. Just like using
Qualtrics, the creator can preview a conversation with the created
chatbot before deploying it (Fig \ref{fig:juji2}). Similar to a
Qualtrics survey, the chatbot is also distributed via a URL.

We chose to use Juji in our study for three reasons. First, the
customization and deployment of a Juji chatbot is very similar to
creating a survey on Qualtrics. This allows survey creators to easily
design, test, and deploy their own conversational surveys, especially
if such chatbot-driven surveys are shown to be valuable. Second, the
Juji platform is publicly accessible, which allows our study to be
easily replicated by others. Third, compared to simple chatbots with
very limited conversational capabilities, the Juji chatbot is equipped
with a rich set of conversational skills
(Table \ref{tab:conversationalfeatures}). Since such conversation
capabilities are considered critical for effective human-agent
interactions \cite{bohus2009ravenclaw}, we are interested in
investigating how such an AI-powered chatbot would affect survey
participants and ultimately the quality of their responses.

Nonetheless, a recent chatbot study finds that it is still quite
challenging to create capable chatbots for handling tasks like meeting
scheduling \cite{grudin2019chatbots}. It is unclear how well the Juji chatbot
would perform in handling more complex tasks such as surveying users
with open-ended questions, where user responses to such questions are
much more diverse and complex to anticipate and handle than those in
meeting scheduling \cite{chen2017effects, devault2014simsensei}. To minimize
potential adverse effects that the Juji chatbot might have on survey
participants and survey results due to its limitations, in this study
we decided to employ the Juji's \textit{all} available interaction
features to maximize its
conversation capabilities (Table \ref{tab:conversationalfeatures}). Accordingly, this study focused on
investigating the holistic effect of the chatbot on survey
participants and survey results, instead of the contribution of individual features, which could be examined in the future with more controlled studies.


We hypothesize that the Juji chatbot's conversational capabilities would
encourage participant engagement and improve the quality of their
responses along several aspects. First, survey research shows that
interactive feedback improves the quality of survey responses
\cite{conrad2005interactive, Bauer2010, Louw2010, kimchi2019}. Since
the Juji chatbot can interactively acknowledge user input including
user free-text input during a survey, this ability may help elicit
quality user responses. Below is a conversation excerpt between the
Juji chatbot and a participant at the beginning of the survey.

\begin{table}[H]
\centering
\begin{tabular}{p{10mm}p{110mm}}
\hline
Juji: & First, let me tell you a bit about myself. I'm an Artificial Intelligence (AI), and I am here to chat with you to learn your thoughts on new games.  \\
Juji: & Now could you tell me about yourself in 2 to 3 sentences? I am curious to know what kind of game player you are. \\
\rowcolor[HTML]{EFEFEF} 
User: & I'm an artist and designer. I put a lot of time and effort into my work. I am extremely driven and competitive, I have lots of goals in life as well.       \\
Juji: & It's a great pleasure to meet someone who is creative. I am very impressed by what you do.  \\ \hline
\end{tabular}
\end{table}

As shown above, Juji's response may make the user feel heard and
thus would be more willing to continue the conversation. However, in a
conversational survey, open-ended questions may elicit arbitrarily
complex or diverse user responses. Below are several example user
responses to the same self-introduction question asked above by the
Juji chatbot.

\vspace{0mm} 

\begin {quote}
    \textit{``I am an artist, I paint, draw, (and) sculpt. I Love to
      cook. my favorite food is pizza. I like playing video games. I wouldn't consider myself to be a hardcore gamer or anything. I love casino games.''}
\end {quote}    

\vspace{0mm}

\begin {quote}
    \textit{``I am a 29 year old single mother that enjoys playing games with my child. It is something we both enjoy!''}
\end {quote}    

\vspace{0mm}

\begin {quote}
\textit{``I am a young adult who plays video games in my spare time. I enjoy video games with compelling story lines that get me invested in characters.''}
\end {quote}

\vspace{0mm}

While the Juji chatbot has a built-in rich dialog library that can
handle a wide range of user input, it may not always be able to
interpret a user input and respond well. Our study thus is also
intended to examine how such a capable but imperfect chatbot would interact with survey participants and influence survey results.

On the one hand, studies show that participants may not always be
cooperative during a survey, which may result in shallow, irrelevant,
or even bogus responses \cite{Egelman2014, muller2014survey}. On the
other hand, research shows that proper ``interventions'', such as
interactive probing and nudging, could help improve response quality
\cite{oudejans2010using, behr2012, deLeeuw2015}. Juji chatbot's
built-in conversational skills provide natural
``interventions''. For example, the Juji chatbot could detect certain
gibberish and irrelevant user responses to an open-ended
question. Below is a chat excerpt between the Juji chatbot and a
participant:

\begin{table}[H]
\centering
\begin{tabular}{p{10mm}p{110mm}}
\hline
Juji: & What do you think of the trailer you just watched?\\
\rowcolor[HTML]{EFEFEF} 
User: & gasaf bafasfasf asfasfas gasfreasfasf afasfa \\
Juji: & Sorry, you must speak a language I don't understand. Could you try English? \\ \hline
\end{tabular}
\end{table}

Some participants might dodge a question by responding with
``\textit{I don't know.}'' or ``\textit{I don't want to answer.}'' In
such cases, the Juji chatbot also has a set of built-in conversation
prompts to handle various user excuses or digressions. Below are two
examples.

\begin{table}[H]
\centering
\begin{tabular}{p{10mm}p{110mm}}
\hline
Juji: & What do you think of the trailer you just watched?\\
\rowcolor[HTML]{EFEFEF} 
User: & no opinion \\
Juji: & I'm pretty sure you have some valuable input. Please do share. \\\hline
\end{tabular}
\end{table}

\begin{table}[H]
\centering
\begin{tabular}{p{10mm}p{110mm}}
\hline
Juji: & What is your favorite game?\\
\rowcolor[HTML]{EFEFEF} 
User: & I don't really have a favorite. \\
Juji: & It doesn't have to be your absolute favorite, just name one you enjoyed or would enjoy.\\\hline
\end{tabular}
\end{table}

Instead of answering a question, a participant may even ask a chatbot
the same question reciprocally. For example, when a participant is asked
``\textit{What do you think of the game trailer you just watched?}'',
the participant might respond: ``\textit{what do you think}''. Again, the Juji chatbot has built-in conversation capabilities
to handle various user questions, including answering a participant's
reciprocal or clarification questions.

To the best of our knowledge, no existing study has examined the use
and effect of chatbots with rich conversational capabilities in the context of conducting surveys with dominantly
open-ended questions. This is also why we chose the Juji chatbot: it has a rich set of conversation capabilities that can handle diverse and complex user free-text input, as well as manage diverse, nonlinear conversation situations \cite{zhoutiis2019}. Moreover, it can be easily customized by feeding it with different survey questions. Equally important, since Juji is a publicly available platform, it should be easy for other researchers and practitioners to
replicate the study presented here to further validate and explore the
values of such a chatbot in facilitating human-subject research.

\subsection {Survey Questions}

Collaborating with the market research firm, we designed a survey that
consisted of mostly open-ended questions with a few choice-based
questions as described below. The survey contained three major parts.

\begin{itemize}

    \item \emph{Warm up}. Each survey started out with 3 open-ended
      questions. A participant was first asked to introduce
      him/herself in 2 to 3 sentences. S/he was then asked to talk
      about his/her favorite games, and what new games s/he is most
      looking forward to playing in the next three months.

      \vspace{2mm} 
      
    \item \emph{Game Trailer Assessment}. Each participant was asked
      to watch two game trailers, one at a time. After watching a
      trailer, the participant was asked to describe his/her thoughts
      and feelings by answering a set of questions:

      \vspace{2mm} 
      
      \begin{itemize}
      \item What is your immediate reaction to this trailer? 
      \item What do you like about it? 
      \item What do you not like about it? 
      \item How interested are you in purchasing the game you just saw in the trailer? Please rate your level of interest in purchasing the game on a scale of 1-5, 1 being no interest, and 5 being very interested. 
      \item Why did you give this score?
      \item How has the trailer influenced your interest?
      \item What would influence your buying decision the most?
      \end {itemize}

      \vspace{0mm} 
In this part, all the questions were open-ended except the rating
question. To avoid potential biases, the order of showing the
two-game trailers was randomly decided for each participant.

\vspace{2mm} 

\item \emph{Additional Information}. Each participant was also asked
  to provide additional information, such as what game platforms they
  use the most, where they look for information about games, and their
  basic demographics including gender, age, and level of education.
    
\end{itemize}

Both the Juji chatbot and Qualtrics surveys used the same set of
questions shown above in the same wording and order. To ensure
consistency, in both conditions participants can take a survey on a
desktop machine, a mobile device, or switch between the two. At the
end of the chatbot survey, the chatbot also asked the participants for
optional comments about their survey experience \footnote{Participants' additional comments were not counted when
measuring \textit{Response Length}}.  

\subsection{Participants}

Our collaborator---the market research firm---hired a panel company to
recruit target participants for the study. The panel company is the
world's second largest company that specializes in recruiting and
managing survey participants for a number of industries. It maintains
a large database of hundreds of millions of survey participants across
all demographics around the world.  In our study, the market research
firm requested the target audience to be US video gamers who are 18
years or older and must play video games at least one hour per
week. Based on these criteria, the panel company queried its database
and found a large pool of candidates whose profiles matched the two
criteria. The pool was randomly divided into two groups where the
Qualtrics link was sent to one group and the chatbot link to another group.

\subsection{Measures}
\label{sec:measures}

To answer our two research questions, we wanted to compare the quality
of collected information (RQ1) and participants' engagement level (RQ2)
between the use of the chatbot survey and the Qualtrics survey. The
survey results were stored in two CSV files, respectively. Each CSV
file contained only question-response pairs. The side talking in a
chatbot survey was not in the CSV file. Instead, such information was
captured in the chat transcripts. Each completed survey was also
stamped with a start and finish time. Most of our analyses shown below
were based on the content captured in the two CSV files.

\subsubsection{Assessing Information Quality}

\hfill \break

Collecting quality information is often the most important goal that a
survey is set out to achieve. Although our surveys contained both
open-ended and choice-based questions, in this study we focused on
assessing and comparing the quality of free-text responses to
open-ended questions for three reasons. First, open-ended questions
were intended to elicit richer and more in-depth input from the
participants, which would enable our collaborator (the marketing
research firm) to better understand gamers' thoughts and feelings and
inform business decisions (e.g., product development and marketing
messages). Second, eliciting quality responses to open-ended questions
has been a major challenge in traditional online surveys because
participants are often not motivated and unwilling to provide
in-depth, quality input. \cite{oudejans2010using}. In this study, we
wish to investigate whether the Juji chatbot's conversational skills,
such as probing and prompting, could help alleviate such a
challenge. Third, it is difficult to determine let alone compare the
quality of user responses to choice-based questions, since it is hard
to tell whether a participant has made a sensible or just a random
choice to such a question during a survey.

To the best of our knowledge, there is no effective tool that can assess the quality of
free-text responses to open-ended questions \emph{automatically}. We
thus had to manually assess the quality of each free-text user
response collected in the surveys. To guide
us to assess the quality of user responses systematically, we
developed a set of content-based metrics based on Gricean Maxims
\cite{grice1975logic}. The Gricean Maxims, proposed by H.P. Grice in
1975, are a set of communication principles to which both speaker and
listener should adhere to ensure effective communication. Gricean
Maxims are often considered “cooperative principles to guide effective
communications” \cite{dybkjaer1996-grice}. In the context of
conducting surveys, a “cooperative” participant would obey all the
maxims to produce quality responses. For example, a participant's
relevant answer to an open-ended question complies with the Gricean
relevance maxim, while a participant's clear response to a question
satifies the Gricean clarity maxim. For our purpose, we use the maxims
to guide us to define a set of metrics that quantitatively measure the quality of information communicated by survey participants. As
shown in Table \ref{tab:measurements}, we measure the quality of
information---a user's response to an open-ended question from four
aspects: \textit{informativeness}, \textit{specificity}, \textit
{relevance}, and \textit{clarity}.

Guided by these quality metrics, two researchers independently went through participants' free-text responses to open-ended questions and manually assessed the quality of each of the
response by the three aspects
(i.e., \emph{relevance}, \emph{specificity}, and \emph{clarity}). More
details about our coding protocol are presented
in \ref{sec:codingprotocol}.

\begin{table}[]
\begin{tabular}{m{15mm}m{30mm}m{20mm}m{60mm}}
\hline
\rowcolor[HTML]{EFEFEF} 
\textbf{Gricean Maxims}             & \textbf{Definition}                                        & \textbf{Our Quality Metrics}                                                                                                & \textbf{Definition}                                                                                                                                           \\ \hline
&&&\\
\multirow{2}{15mm}{\textbf{Quantity}}& \multirow{2}{30mm}{One should be as informative as possible} & \textit{Informativeness}                                                                                                    & A participant's response should be as informative as possible                                                                                                 \\
& & \textit{Specificity}                                                                                                        & A response should give as much information as needed.                                                                                                      \\
&&&\\
\rowcolor[HTML]{EFEFEF} 
&&&\\
\rowcolor[HTML]{EFEFEF} 
\textbf{Quality}                    & One should communicate truthfully                          & \multicolumn{2}{m{83mm}}{\cellcolor[HTML]{EFEFEF}A participant's response should be authentic. Since it is difficult to assess the truthfulness of a user input, we didn't measure this aspect directly. Our another measure, the level of self-disclosure might signal a level of authenticity.} \\
\rowcolor[HTML]{EFEFEF} 
&&&\\
&&&\\
\textbf{Relevance}                  & One should provide relevant information                    & \textit{Relevance}                                                                                                          & A participant's response should be relevant to a question asked                                                                                               \\
&&&\\
\rowcolor[HTML]{EFEFEF} 
&&&\\
\rowcolor[HTML]{EFEFEF} 
\textbf{Manner}                     & One should communicate in a clear and orderly manner       & \textit{Clarity}                                                                                                            & A participant's response should be clear                                                                                                                      \\
\rowcolor[HTML]{EFEFEF} 
&&&\\\hline
&&&\\
\end{tabular}
\caption{Gricean Maxims used to guide the development of information quality metrics.}
\label{tab:measurements}
\end{table}

\paragraph{Informativeness}
By the Gricean Maxim of quantity, an effective communication should be informative. To measure the \emph{informativeness} of a text response, we computed the amount of information conveyed in the response by \emph{bits} (shannons) based on information theory \cite{jones1979elementary}. More precisely, the informativeness of a text response is the sum of each of its word's \textit{surprisal}, the inverse of its expected frequency appearing in modern English (Formula \ref{eq:info}). In other words, the more frequently a word (e.g., the common word ``the'') appears in modern English communications, the less information it conveys.

\begin{equation}
I(Response) = \sum\log_2\frac{1}{F(word_n)}
\label{eq:info}
\end{equation}
\vspace{0mm}

To obtain an accurate estimate of a word's frequency in modern English, we averaged a word's frequencies in four text corpora, the British National Corpus \cite{leech1992100}, The Brown Corpus \cite{hofland1982word}, Webtext \cite{parviainen_2010}, and the NPS Chat Corpus \cite{forsythand2007lexical}. For each participant, we computed a total \textit{informativeness} based on his/her free-text responses to all open-ended questions.

\paragraph{Specificity}
Although our informativeness metric mentioned above computes the
amount of information conveyed by a user's text response, it does not
assess how specific the response is.  Specific responses often provide
sufficient details, which not only help information collectors better
understand and utilize the responses, but also enable them to acquire
more valuable, in-depth insights. For a given open-ended question,
text responses could be very diverse, complex or even ambiguous. Since
we could not find a reliable natural language processing tool to
assess the specificity of diverse text responses to a given question
automatically, we manually assessed the \textit{specificity} of each
text response on three levels: 0 - generic description only, 1 -
specific concepts, 2 - specific concepts with detailed examples.

By our specificity metric, a response would obtain a level-0
specificity if it provides only a shallow or abstract description. For
example, when asked ``\emph{What is your immediate reaction to this
trailer?}'', a typical shallow response with level-0 specificity was

\vspace{0mm} 

\begin {quote}
\textit{``I love it, it looks interesting.''}
\end {quote}

\vspace{0mm} 

In contrast, a response with level-1 specificity conveys more
specific information, such as the following statement:

\vspace{0mm} 

\begin {quote}
\textit{``I am interested in the game and I really like the graphic''}.
\end {quote}

\vspace{0mm} 

The most specific responses with a specificity score of 2 normally offer
detailed descriptions. For example, one such response stated

\vspace{0mm} 

\begin {quote}
\textit{``Completely blown away! It is unlike any game I have ever
  seen! There are so many different scenes, so many different
  characters who look different from each other, a lot of cool weapons
  and gadgets, so many different ways of fighting, so many different
  places you can go and it is so action-packed''}
\end {quote}

\vspace{0mm}

\paragraph{Relevance}

By the Gricean Maxim of relevance, a quality communication should be
relevant to the communication context. In a survey context, a quality
response should be relevant to the survey question asked. Not only do
irrelevant responses provide no value, but they also burden the
analysis process. For a given open-ended question, text responses
could be very diverse and complex. Similar to assessing
the \textit{specificity} of a text response, we manually assessed
the \textit{relevance} of each text response on three levels: 0 -
Irrelevant, 1 - Somewhat Relevant, and 2 - Relevant.

A response was considered irrelevant if it did not relate to the
question asked at all. For example, a gibberish response like
\emph{``Yhhchxbxb''} was considered irrelevant and received a relevance
rating 0. Certain responses were considered partially relevant as they
did not answer an asked question directly but still provided useful
input. Here is an example response that was assigned a relevance score
1, when asked \textit{``What do you like about it (the game
trailer)?''}, a participant responded:

\vspace{0mm}

\begin {quote}
\textit{``I don't like it. I don't purchase these types of games. But for people
  who enjoy these types, they might enjoy it.''}
\end {quote}

\vspace{0mm}

Responses that directly and clearly answered an asked question were considered relevant and assigned a relevance score 2.

For each participant, a total relevance score was also computed by adding up the relevance scores of each response.

\paragraph{Clarity}
By the Gricean Maxim of clarity, an effective communication should be the \textit{clarity} of each text response by how easily the response could be understood by humans without ambiguity, regardless of its topical focus, on three levels: 0 - illegible text, 1 - incomplete sentences, 2-clearly articulated response. Again, given today's natural language processing capabilities, automatically and reliably scoring the \textit{clarity} of a text response is difficult. We decided to manually score the clarify of each text response.

Gibberish or nonsense responses were marked with 0. We marked responses as
partially legible with a score of 1 if they contained incomplete
sentences or grammatical errors that impeded a reader's ability to
interpret the responses. For example, when asked \textit{``What is
your immediate reaction to the trailer''}, a response \textit{``very
good''} was marked with a clarity score of 1. Responses obtained a
clarity score of 2, if they were articulated with completed sentences
with no serious grammatical issues.

\subsubsection{Measuring Level of Engagement}

In a typical online survey setting, evidence shows that participants do not tolerate long surveys. As a survey grows in length, the time spent on each question dramatically decreases, and the completion rate also drops significantly. The tolerance for lengthier surveys is even lower for customer-related surveys like the one in our study.

Because a chatbot survey is still a novelty but with flaws (e.g., unable to handle certain user input), we wanted to find out how it would
impact participant engagement. In particular, we measured the level of \textit{participant engagement} from three aspects: \textit{engagement duration}, \textit{response length}, and \textit{self-disclosure}.

\paragraph{Engagement Duration}

\emph{Engagement duration} measures how long a participant takes to complete a survey. A longer engagement duration suggests that a participant more willingly stays engaged longer. This was especially true in our case since each participant was rewarded by completing a survey, and not how much time s/he took. For each participant, the engagement duration was automatically logged by each platform.

\paragraph{Response Length}
\emph{Response length} is the word count of each participant's free-text responses. Similar to engagement duration, the response length also signals participants' willingness to stay engaged if they are willing to write longer responses.

\paragraph{Self-Disclosure}

Self-disclosure is often used as an indicator for measuring
human-agent engagement, as reflected by the breadth and depth of
topics exchanged in human-computer conversations \cite{bickmore2011relational,shamekhi2018face}, based on the social
penetration theory \cite{altman1973social}. Self-disclosure is
particularly important for survey research that aims to elicit
personal thoughts and feelings. To measure \textit{self-disclosure},
we manually analyzed each participant's response to the
self-introduction question, and manually counted the number of
attributes or topics mentioned (e.g., age, gender, and hobbies). Below
lists several example participant responses that were coded with
varied level of self-disclosure.

Participants who were most willing to disclose about themselves often offered
detailed descriptions about themselves in their responses. For
example, the following response mentions the participant's \emph{age},
\emph{gender}, \emph{marital status}, \emph{favorite game type}, \emph{favorite game}, \emph{game playing history} and even \emph{living condition}.

\vspace{0mm}
\begin {quote}
    \textit{``I'm a 29 year old single guy living alone adn i love RPG
    I really like games like fortnite and call of duty...those have
    always been my most favorite games and I continue to like them
    into adult hood...i lose track of time playing them''}
\end {quote}  
\vspace{0mm}
In contrast, the response below said much less about the participant
except the video game s/he plays.
\vspace{0mm}
\begin {quote}
    \textit{``The only video games I play is mariokart.''}
\end {quote}  

Since Gricean Maxims serve merely as theoretical guidelines for us to
measure the quality of user free-text responses to open-ended
questions, our current metrics are just one of many ways to
\emph{estimate} the quality of user free-text responses. By no means are these metrics 
unique. Other similar metrics may be defined based on specific
situations. For example, if a survey cares more about getting relevant
responses than the specificity of the responses, different weights
might be associated with each aspect to compute a
weighted \emph{response quality index} (RQI). In general, Gricean
Maxims can be used as a framework to guide the definition of
computational metrics that measure communication quality. The
framework allows researchers who are interested in measuring survey response quality to use a systematic method to do so. 

\subsubsection{Coding Protocol}
\label{sec:codingprotocol}

Many of the metrics mentioned above require human coding
effort. Since we collected over 11,000 free-text responses to nineteen
open-ended questions, manually coding each response on all the
metrics would require tremendous effort \footnote{We did consider
the use of Amazon Mechanical Turkers to help code the data. But we
could not do so due to the confidentiality of the results.}. We thus
worked with our collaborator---the market research firm to first
identify the analysis requirements based on their business needs. Per
the purpose of the study, they selected the top nine most important
questions of which answers would help them derive the desired market
insights. We manually analyzed a total of 5238 text responses to these
nine questions and coded each response on
its \emph{relevance}, \emph{clarity},
and \emph{specificity}. We manually assessed 582
participant responses to the self-introduction question and coded
each response on its level of \emph{self-disclosure}.

In general, we used a 3-step process to manually code the text
responses. First, two human coders independently rated about 10\% of
randomly sampled responses of each selected question. Second, they
reconciled differences and came up with a set of more consistent coding
criteria. Third, they used the established coding criteria to code
the rest of the responses independently. To avoid potential biases,
the coders were blind to the source of responses. After all responses
were coded, a Krippendorff's alpha was used to measure the
inter-rater reliability of the coded results
\cite{krippendorff2011computing}. If the Krippendorff's alpha was
above 0.8, indicating a high level of agreement between the coders,
the coders then moved on to reconcile the remaining differences if
there were any. Otherwise, the coders discussed the differences and
re-iterated on the coding process. The Krippendorff's alpha ranged
from 0.80 to 0.99 for each set of coding.

\section{Results}
We first provide an overview of our results followed by detailed analyses.
\vspace{-2mm}
\subsection{Overview}
\label{sec:overview}

\begin{figure}[t]
\centering
\includegraphics[width = 9cm]{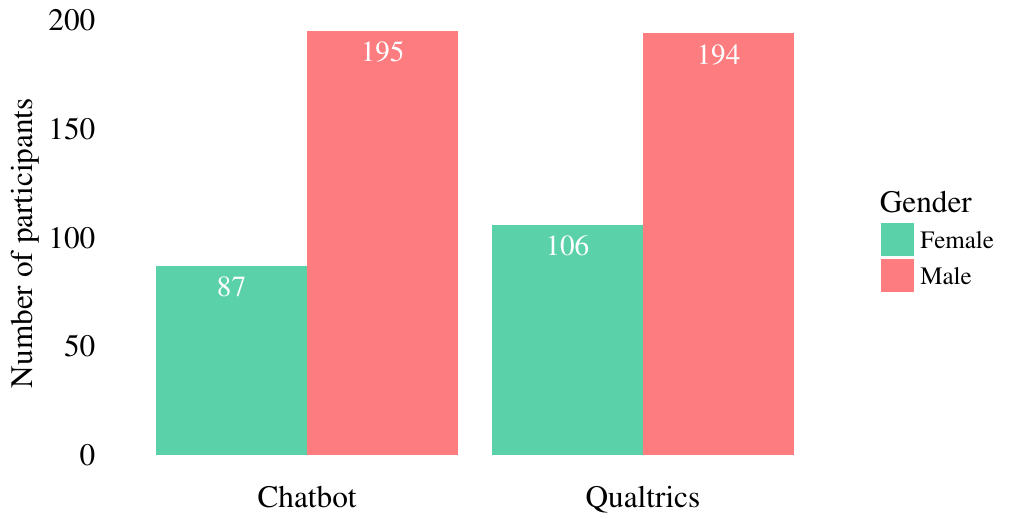}
\caption{Distribution of Participants by Gender.}
\label{fig:gender}
\end{figure}

\begin{figure}[t]
  \begin{subfigure}[b]{0.39\textwidth}
    \centering
    \includegraphics[width=\textwidth]{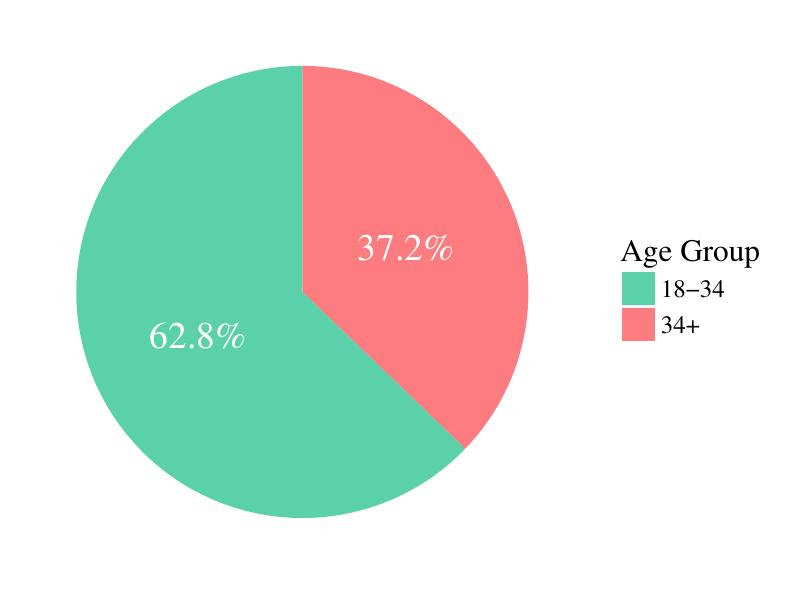}
    \caption{Chatbot}
    \label{fig:1}
  \end{subfigure}
  \begin{subfigure}[b]{0.39\textwidth}
    \centering
    \includegraphics[width=\textwidth]{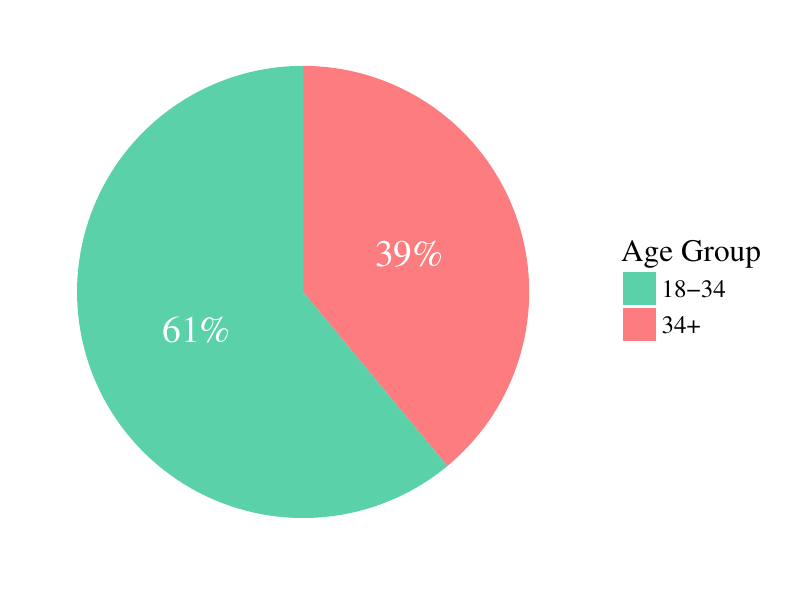}
    \caption{Qualtrics}
    \label{fig:2}
  \end{subfigure}
  \caption{Age Distribution of Participants.}
  \label{fig:age}
\end{figure}

\begin{figure}[]
  \begin{subfigure}[b]{0.49\textwidth}
      \centering
    \includegraphics[width=\textwidth]{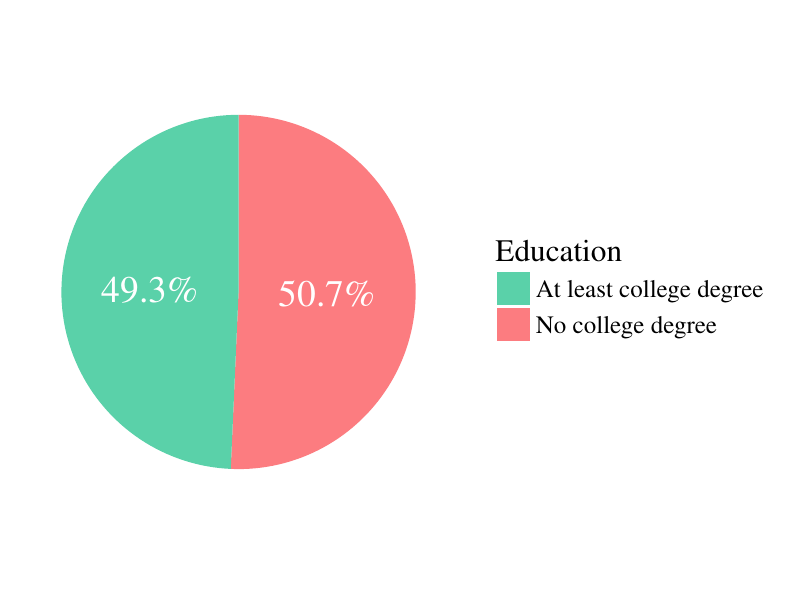}
    \caption{Chatbot}
    \label{fig:1}
  \end{subfigure}
  \begin{subfigure}[b]{0.49\textwidth}
      \centering
    \includegraphics[width=\textwidth]{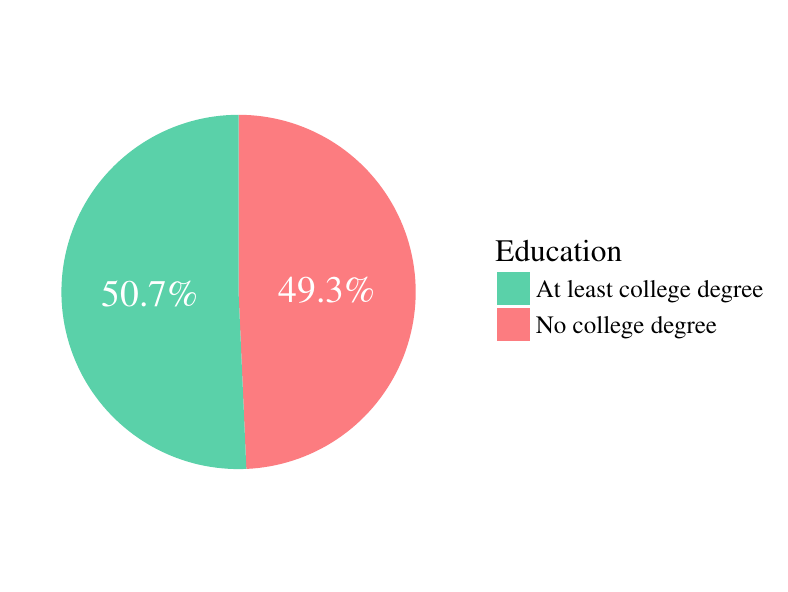}
    \caption{Qualtrics}
    \label{fig:2}
  \end{subfigure}
  \caption{Distribution of Participants by Education Level.}
  \label{fig:edu}
\end{figure}

We received a total of 582 completed surveys: 282 chatbot surveys and
300 Qualtrics surveys. As shown in Fig \ref{fig:gender}, among the
282 chatbot survey takers, 87 (30.85\%) were female, and 195 (69.15\%)
were male; while 106 (35.33\%) of 300 Qualtrics participants were
female, and 194 (64.67\%) were male. Fig \ref{fig:age} and Fig \ref{fig:edu} show the distribution of participants by their age
group and level of education. The participants' ages ranged from 18-50
years old, where the majority (61.86\%) of them were between 18-34
years old. Among the 582 people who completed their survey, 50\%
received at least a college degree. The average self-reported weekly
gaming time is 16.90 hours (SD = 13.50 hours).

To measure survey success, two key metrics are often used: \emph{response
rate} (Equation \ref{fun:response}) and \emph{completion rate}
(Equation \ref{fun:completion}) \cite{Rossibook2013}.

\begin{equation}
   \text{response rate} = \frac{\text{number of participants clicked on a survey link}}{\text{number of the participants invited}}
   \label{fun:response}
\end{equation} 

\begin{equation}
   \text{completion rate} = \frac{\text{number of participants completed a survey}}{\text{number of participants clicked on a survey link and qualified}}
    \label{fun:completion}
\end{equation}

\begin{figure}[t]
\centering
\includegraphics[width = 9cm]{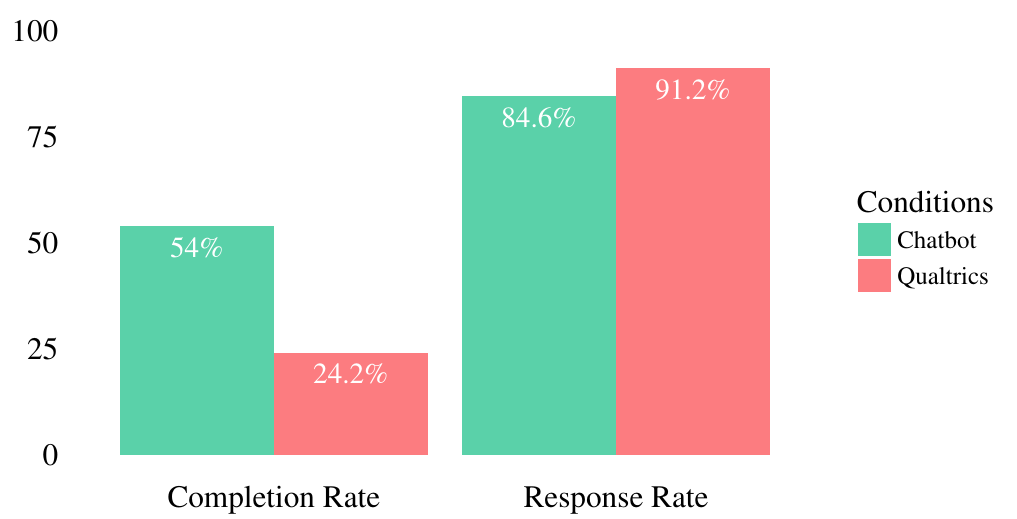}
\caption{The response rate and completion rate of the Chatbot survey and the Qualitrics survey}
\label{fig:response_completion}
\end{figure}
\vspace{0mm}

Based on the data provided by the panel company, Fig \ref{fig:response_completion} shows the \emph{response rate} and
\emph{completion rate} for the chatbot and Qualtrics surveys,
respectively. The \emph{response rate}
of the chatbot group (84.6\%) was lower than that of the Qualtrics
group (91.2\%). A Two Proportion Z-test shows that the difference is significant: z = 4.35, p < 0.01**. The difference may be due to participant's unfamiliarity to the conversational survey (see Sec \ref{sec:studycontrols} for further discussion). However, the
\emph{completion rate} of the chatbot group (54\%) was more than twice
(2.2 times) the \emph{completion rate} of the Qualtrics group (24.2\%).
A Two Proportion Z-test shows that the difference is significant: z = -12.16, p < 0.01**. The higher completion rate suggests the
better ``stickiness'' of the chatbot survey.


Below we present the results to answer our two research
questions. Since the goal of our study is to compare the outcomes of
two survey methods, chatbot versus Qualtrics, on two sets of measures
(response quality metrics and participant engagement metrics), we
chose to run ANCOVA analyses. ANCOVA is a general linear model
blending ANOVA and regression, which helps us examine the true effect
of the survey method \cite{aaker1976design}. In each ANCOVA analysis
shown below, the independent variable was the survey method used, and
the dependent variable was a computed response quality metric or an
engagement metric. Since research suggests that demographics influence
people's behavior with new technology \cite{venkatesh2000longitudinal, morris2000age}, all analyses was
controlled for participants' age, gender, education level and weekly gaming time. Each analysis
was also controlled for by participants' weekly gaming time, as research
shows that game playing experience impacts people's beliefs and
attitude toward technology \cite{hayes2013gamer,
  Carstens2004}. Furthermore, the analyses involving \emph{response quality},
\emph{response length}, and \emph{self-disclosure} were also
controlled for \emph{engagement duration}, since we wished to assess
whether the chatbot led to improved information quality, response
length, and self-disclosure even when controlled for the amount of time
that participants spent on completing a survey.

Before running ANCOVA analyses, we first examined the correlations
among all our dependent variables (i.e., response quality and
participant engagement metrics) to see how they may be related to each
other (Table \ref{tab:measure correlation}). It is interesting to
note that most of the variables were correlated except
\emph{engagement duration}, which did not significantly correlate with any other
metrics except \emph{response length}. No interaction effects were found. Intuitively, this result seems
sensible since most of the metrics were assessing the \emph{content}
of participants' responses (e.g., \emph{specificity} and \emph{self-disclosure}). This also implies that \emph{engagement duration}
\emph{alone} would not signal the quality of participant responses.

\begin{table}[t]
\begin{threeparttable}
\begin{tabular}{llllllll}
\hline
Outcome Measures       & 1      & 2      & 3    & 4    & 5    & 6    & 7  \\ \hline
\cellcolor[HTML]{EFEFEF}Response Quality    &               &              &                &               &                     &                     &                                \\
1. Informativeness     & --     &        &      &      &      &      &    \\
2. Relevance           & 0.36** & --     &      &      &      &      &    \\
3. Clarity             & 0.45** & 0.92** & --   &      &      &      &    \\
4. Specificity         & 0.60** & 0.75** & 0.80** & --   &      &      &    \\
\cellcolor[HTML]{EFEFEF}Participant Engagement    &               &              &                &               &                     &                     &                                \\
5. Engagement Duration & 0.06   & 0.04   & 0.02 & 0.03 & --   &      &    \\
6. Response Length     & 0.84** & 0.31** & 0.41** & 0.55** & 0.10** & --   &    \\
7. Self-Disclosure     & 0.11** & 0.18** & 0.19** & 0.25** & 0.03 & 0.13** & -- \\ \hline
\end{tabular}
\begin{tablenotes}
    \item[a] N = 582.
    \item[b] *p < 0.05, ** p < 0.01. *** p < 0.001
\end{tablenotes}
\caption{Correlations between dependent measures}
\label{tab:measure correlation}
\end{threeparttable}
\end{table}

Table \ref{tab:resultsummary} summarizes the ANCOVA analysis results
for each measure, of which details will be discussed below.

\subsection{RQ1: How Would the Quality of Responses Differ?}

As mentioned in Section \ref{sec:Method}, we have developed a set of metrics to
measure the quality of a user response from four aspects in
Table \ref{tab:measurements}. Using the responses collected by the two survey
methods, we compared their quality by each metric.

\begin{table*}[t]
\setlength\belowcaptionskip{10pt}
\begin{threeparttable}
\begin{tabularx}{\linewidth}{l|ll|ll|l|l|l}
                                               & \multicolumn{2}{l|}{Chatbot} & \multicolumn{2}{l|}{Qualtrics} &                     &                     &                                \\ \cline{2-5}
\multirow{-2}{*}{Measures}                     & M             & SD           & M              & SD            & \multirow{-2}{*}{F} & \multirow{-2}{*}{p} & \multirow{-2}{*}{$\eta_{p}^2$} \\ \hline
\cellcolor[HTML]{EFEFEF}Responses Quality    &               &              &                &               &                     &                     &                                \\
Informativeness (bits)                      & 283.33        & 152.90       & 203.53         & 184.19        & F(1, 576)=38.55               & <0.01**   & 0.06                           \\
Relevance                                      & 15.72         & 4.16         & 14.05          & 5.55          & F(1, 576)=17.63               & <0.01**   & 0.03                           \\
Response Quality Index                         & 27.28         & 10.20        & 21.70          & 10.31         & F(1, 576)=48.72               & <0.01**   & 0.08                           \\ \hline
\cellcolor[HTML]{EFEFEF}Participant Engagement &               &              &                &               &                     &                     &                                \\
Engagement Duration (mins)                     & 24.38         & 13.42        & 17.90          & 17.20         & F(1, 576)=24.60               & <0.01**   & 0.03                           \\
Response Length (words)                        & 90.11         & 46.23        & 63.98          & 54.17         & F(1, 576)=57.92               & <0.01**   & 0.09                           \\
Self-Disclosure                                & 5.16          & 2.26         & 3.57           & 2.45          & F(1, 576)=34.82               & <0.01**   & 0.06                          

\end{tabularx}
\begin{tablenotes}
    \item[a] All results were controlled for participant's demographics, including gender, age, education level, and weekly gaming time.
    \item[b] Results for Responses Quality (including Informativeness, Relevance, and Response Quality Index), Response Length and Self-Disclosure were additionally controlled for participant's Engagement Duration.
    \item[c] Results for Engagement Duration have additional control for Response Length.
\end{tablenotes}
\caption{Results summary including ANCOVA analysis results on individual measures}
\label{tab:resultsummary}
\end{threeparttable}
\end{table*}

\subsubsection{Informativeness}

By Formula \ref{eq:info}, we computed an \emph{informativeness} score
of each completed survey based on the participant responses given in
that survey. Our results showed that on average the chatbot surveys
collected 39\% more information than the Qualtrics surveys. With the
survey method as its independent variable and controlling for
demographics (i.e., gender, age, and education level), weekly
game-playing time, and engagement duration, an ANCOVA analysis Table \ref{tab:resultsummary} showed
that the chatbot surveys collected significantly richer information
than the Qualtrics surveys, and the survey method was a significant
factor contributing to such differences. In addition, among the
control variables, the level of education, was the only factor shown
to be significant, although there was no interaction effect between
the survey method and the educational level. Specifically,
participants with at least a college degree (M = 259.93 bits, SD =
166.14 bits) offered richer responses than those without a college
degree (M = 224.46 bits, SD = 180.53 bits): F (1, 576) = 6.81, p <
0.01**, $\eta_{p}^2$ = 0.01. There was no evidence suggesting the
effect of age, gender, engagement duration, or game-playing time.

\subsubsection{Relevance}

Next, we examined the \emph{relevance} of collected responses. As
mentioned in Section 3.5, we manually assessed the \emph{relevance} of
participants' free-text responses to a selected set of nine open-ended
questions. For each completed survey, we created a \emph{relevance
  index} by combining all its responses' relevance scores
additively. The results showed that on average the chatbot surveys
collected 12\% more relevant responses than the Qualtrics surveys did.

With the survey method as the independent variable and controlling for
demographics, game-playing time, and engagement duration, an ANCOVA
analysis Table \ref{tab:resultsummary} revealed that the survey method contributed to the
differences in \emph{relevance} significantly. In other words, the
participants who completed a chatbot survey provided more relevant
responses than those who finished a Qualtrics survey. Also results
showed that people who played more games per week tended to provide
more relevant responses in a survey ($\beta = 0.04$, p < 0.05*). This
result suggests that enthusiastic gamers perhaps are more receptive to
chatbots and more willing to offer quality information during their
interaction with the chatbots. No interaction effects were found.

To help us better understand the differences in response
\emph{relevance}, we further examined the surveys with a
\emph{relevance index} value of zero (0), which implied none of their
responses was relevant. We found that 27 (9.00\%) out of 300 completed
Qualtrics surveys contained all gibberish (e.g., \emph{"fdlfdbdffdh"}
or its variants) or bogus statements (e.g., \emph{"Funding from a
  state Itsdhzxoy"} given as a self-intro). In contrast, only 7
(2.48\%) out of 282 completed chatbot surveys contained completely
irrelevant responses\footnote{Although the Juji chatbot can detect
  certain gibberish, its gibberish detection was turned off for
  certain questions. For example, it was turned off for questions asking about one's favorite game or game platform. This is because many legit game names or game platforms might be considered gibberish as they don't exist in the generic natural language corpora used for Juji's gibberish detection algorithm.}. A Two Proportion Z-test showed a significant difference in the proportion of gibberish responses between two conditions (z = 3.35, p < 0.01**). This also
implies that participants were less likely to ``cheat'' when interacting
with a chatbot in a survey.  Due to inadequate data collected (see
``Study Limitations'' under Section 5), it is however unclear which
chatbot behavior contributed to the result (e.g., probing, prompting,
and social commenting). It might be the combination of the perceived
anthropomorphic characteristics of the chatbot and the novelty
factor. From the participants' own comments left at the end of a
chatbot survey, it is the fact that most of the participants have not
experienced a chatbot-driven conversational survey and they were not
fully aware of the chatbot's capabilities yet.

\subsubsection{Response Quality Index}

For each relevant response, we further examined its quality by two more metrics: \emph{specificity} and \emph{clarity}. To do so, we created an overall \emph{response quality index (RQI)} by aggregating the three quality metrics:

\begin{equation}
\begin{array}{l}
   RQI = \sum_{n=1}^{N} relevance[i] \times clarity[i] \times specificity[i] \\ 
   \text{(N is the number of responses in a completed survey)}
   \end{array}
\label{equ:rqi}
\end{equation} 

By the above formula, we computed the \emph{overall response quality}
of each completed survey. The results showed that on average the
completed chatbot surveys produced 25.7\% better quality responses
than the Qualtrics surveys did. Using the \emph{RQI} as an independent
variable and controlling for demographics, game-playing time, and
engagement duration, an ANCOVA analysis again showed that the survey
method was a significant factor that contributed to the quality
differences. Not only did the participants who completed a chatbot
survey offer more relevant responses, but their responses were also
more specific and clearer than those collected by the Qualtrics
surveys.

In particular, the participants in the chatbot survey often offered
more specific details per the chatbot's question. Here is an example
response to the question \emph{``what's your immediate reaction to the
  trailer''}:

\begin{quote}
\textit{``amazing and very fluid. I like the pace of the game, the
  weapons, the soldiers gears as well as the setting or the terrains
  where the action take place.''} 
\end{quote}

Even a short response in the chatbot survey contained specifics. Below
is such a short response to the same question above:
\begin{quote}
\textit{``I like the scenery. It shows old, broken down building.''} 
\end{quote}

In contrast, the responses collected by the Qualtrics surveys were
more terse and abstract. Here are some example responses to the same
question above

\begin{quote}
\textit{``It looks interesting.''}
\end{quote}

\begin{quote}
\textit{``very good''}
\end{quote}

By the response quality index (\emph{RQI}), participants with at least
a college degree (M = 23.46, SD = 11.79) also provided higher quality
responses than those without (M = 19.89, SD = 12.90) ; F(1, 576) =
14.31, p < 0.01**, $\eta_{p}^2$ = 0.02. Intuitively, this result seems
sensible since the level of education would influence one's
knowledge and communication.

\subsection{RQ2: How Would a Chatbot Impact Participant Engagement?}

To compare participant engagement exhibited in the two survey methods,
we examined three measures: \emph{engagement duration}, \emph{response
  length}, and the level of \emph{self-disclosure}.

\subsubsection{Engagement Duration}
For a completed survey, \emph{engagement duration} recorded how much
time a participant took to finish the survey. Our result showed that
participants took seven more minutes on average to complete a chatbot
survey than finishing a Qualtrics survey. With the survey method as an
independent variable and controlling for demographics, game-playing
time and response length, an ANCOVA analysis showed that the duration
differences were significant, and the survey method was the only
significant factor contributing to such differences. 

Coupling with the survey completion rate (chatbot surveys 54\%
vs. Qualtrics 24\%), the result implied that the participants were
willingly engaged with the Juji chatbot longer. This is especially
true considering that the participants were paid just for completing a survey not for the amount of time spent.

\subsubsection{Response Length}

A longer engagement duration does not necessarily mean that a
participant is willing to contribute more content. We thus measured
\emph{response length} to estimate a participant's level of engagement
by his/her willingness to make content contributions. We counted the
number of words that participants contributed to each completed
survey. On average, the participants contributed 30 more words in a
chatbot survey than in a Qualtrics survey. Using \emph{response
  length} as an independent variable controlling for demographics,
gaming time, and engagement duration, an ANCOVA analysis showed that
such differences were significant and the survey method was the only
significant factor contributing to the differences (Table \ref{tab:resultsummary}).

This result implies that the participants were also willing to contribute
more content in a chatbot survey regardless of their demographics,
game-playing time, and the time spent with a chatbot.

\subsubsection{Self-disclosure}

Research shows that one's willingness to disclose him/herself in a
human-to-human or human-to-machine interaction indicates one's level
of engagement in the interaction \cite{schulman2009persuading,bickmore2010maintaining,bickmore2011relational}. In
our study, since each participant was asked to introduce him/herself
(\emph{``Could you describe yourself in 2-3 sentences''}), we examined
how much information the participant disclosed about him/herself.  Our
coded results showed that on average the participants revealed 1.6
more types of information about themselves (e.g., age, occupation,
pet, and game preferences) in a chatbot survey than in a Qualtrics
survey. An ANCOVA analysis, controlled for demographics, game-playing
time, and engagement duration, showed that such differences were
significant and the survey method was the only significant factor
contributing to such differences.

We further examined the types of information disclosed by the
participants. We found that 32.62\% of the participants disclosed
personal information (e.g., age, gender, and marital status) to the
chatbot, while only 15.67\% of the participants offered similar
information in the Qualtrics survey. Specifically, the participants
disclosed various types of detailed information about themselves in a
chat, such as personal facts, daily activities, and personality. Below
are three example responses from the chatbot surveys.
\vspace{0mm}
\begin{quote}
\textit{``Well I am a student working towards my masters at the same
  time working part-time at Starbucks on campus. I study/work in the
  morning and game at night Spending 8 hours studying/working , 5
  napping, and the rest gaming. Snacking through out the whole time
  but I try to get at least one hour every day of my game time
  exercising.''}
\end{quote}
\vspace{0mm}
\begin{quote}
  \textit{``I am 44 years old, married to my soulmate with a dog named
    Killer.  I like to play guitar and make my own music. I enjoy
    computers and anything techie. I work part time and make money
    online doing different things.''}
\end{quote}
\vspace{0mm}
\begin{quote}
  \textit{``I am a chill laid back person. I like sports and music.''}
\end{quote}
\vspace{0mm}

In contrast, much fewer participants did so in their Qualtrics
surveys. Their responses were mostly about the type of games they
like. Below is an example.

\begin{quote}
  \textit{``I like to play games like creating kingdoms and restaurants.''}
\end{quote}

Asking a participant to make a self-introduction was the very first
question in the chatbot survey after the chatbot said hello and
introduced itself \footnote{The Juji chatbot has a default conversation
  opening that can be customized by a creator}. We thus conjecture
that participants' willingness to self disclose could be attributed to
a tendency of reciprocity, which was found to deepen users' self-disclosure in previous studies of human-agent interaction \cite{moon1998intimate, bickmore2010maintaining}.

\subsubsection{Participants' Feedback}

Quantitatively, our analyses show the significant differences in both
response quality and participant engagement resulted from two types of
surveys and in most cases the survey method is the only significant
factor contributing to such differences. To better understand how the
participants felt about their experience with the chatbot, we also
examined their voluntary comments left at the end of each chatbot
survey. 

Among the 282 participants who finished a chatbot survey, 70\% (193
out of 277\footnote{5 participants left gibberish comments and were
discounted}) left optional comments at the end of their survey. 95\%
(183 out of 193) were positive, 2\% (3 out 193) were negative, and the
remainder were neutral \footnote{The most of the neutral comments
simply said \emph{``thank you''}}. The positive comments roughly fell
into four categories. Among the positive comments, 30\% (54 out of 183
positive comments) indicated personal connections with the Juji
chatbot, and 43\% (79 out of 183) positive comments expressed how much
the participants enjoyed the chat and found the survey
entertaining. In addition, 22\% (40 out of 183) praised that chatting
with Juji was the best survey format that they had experienced, while
the remaining 5\% commented how smart the chatbot was. In short, based
on their comments 67\% of the participants enjoyed their engagement
with the chatbot for one of the main reasons identified above. Below
shows a list of example comments.

One third of participants seemed to have made a personal connection
with the chatbot.
\vspace{0mm}
\begin{quote}
\textit{``You are my new best friend Juji!''}
\end{quote}
\vspace{0mm}
\begin{quote}
\textit{``the whole time i was doing this survey it felt like i was talking to a friend and sharing the same common ground. i loved that i wish it didnt have to end"}
\end{quote}
\vspace{0mm}
\begin{quote}
\textit{``you were great, Juji ... i love you''}
\end{quote}
\vspace{0mm}

43\% of participants simply enjoyed their chat with Juji and had a good time.
\vspace{0mm}
\begin{quote}
\textit{``I have enjoyed our chat and hope you have a great 4th of July''}
\end{quote}
\vspace{0mm}
\begin{quote}
\textit{``I had an amazing time! Juji is AWESOME and darn charming!''}
\end{quote}
\vspace{0mm}
\begin{quote}
\textit{``very dynamic and very fluid conversation you have great quality thanks''}
\end{quote}
\vspace{0mm}
Some thought the chatbot is super smart and cool.
\vspace{0mm}
\begin{quote}
\textit{``You're cool, bro''}
\end{quote}
\vspace{0mm}
\begin{quote}
\textit{``are u really a robot''}
\end{quote}
\vspace{0mm}

About a quarter of participants considered the chatbot survey was fun
and the best survey experience they ever had.
\vspace{0mm}
\begin{quote}
\textit{``This was the most fun I've had ever taking a survey, I
  absolutely loved it. Please do more in the future, it was a fresh
  experience!''}
\end{quote}
\vspace{0mm}
\begin{quote}
\textit{``This was one of the most entertaining surveys ive done. Great Job at making it not boring like others.''}
\end{quote}
\vspace{0mm}
\begin{quote}
\textit{``I have been doing surveys for years now, and this by far is probably the best format that 
I have ever seen, this actually was a very good experience.''}
\end{quote}
\vspace{0mm}
\begin{quote}
\textit{"i like this type of survey experience it makes it more personal and fun Thanks for chatting Juji"}
\end{quote}
\vspace{0mm}

Among just a couple of negative comments, one was complaining about
the reward s/he would receive for taking this survey.

\vspace{0mm}
\begin{quote}
  \textit{``the gift sucked, thought it would be something gaming related or a small gift card''}
\end{quote}
\vspace{0mm}
The other negative comment stated:
\vspace{0mm}
\begin{quote}
  \textit{``I would prefer to not have an ai for a survey''} \end{quote}
\vspace{0mm}

\subsection{Summary of Findings}

As captured in part in Table \ref{tab:resultsummary}, our study
results revealed three key findings as summarized below.
\begin{itemize}
    \item \emph{\textbf{The chatbot elicited significantly higher quality
        responses}}. The participants who completed a chatbot survey
      offered much more relevant, specific, and clear responses to
      \emph{open-ended} questions than their counterparts did in a
      Qualtrics survey.
    
    \vspace{2mm}
    
    \item \emph{\textbf{The chatbot encouraged significantly more
        participant engagement}}. The participants were willingly
      spending more time with the Juji chatbot, writing longer
      responses, and disclosing more information about themselves in
      depth and scope.
    \vspace{2mm}
    \item \emph{\textbf{The chatbot pleased a majority of participants}}.
      The participants' comments show that they enjoyed chatting with
      Juji and preferred taking this kind of conversational survey
      in the future. Even considering the novelty factor and the
      potential positivity tendency toward a humanized machine \cite{thomaz2008teachable}, these comments were still encouraging especially considering that the comments were from
      seasoned, paid survey takers and there was no additional reward
      for them to spend more time on a survey or leave optional
      comments.
\end{itemize}
\section{Discussion}
Here we discuss the benefits and risks of
chatbot-driven surveys, the limitations of our study, and design implications of creating effective
chatbots for conducting surveys.

\subsection {Benefits and Risks of Chatbot Surveys}

The main purpose of surveys is to elicit quality information from a
target audience to inform decisions. One of the most effective ways to
elicit quality information is through an engaging conversation
\cite{Louw2010}. However, having a conversation, especially one
with a human interviewer, may be time-consuming or induce potential biases. In
addition to the findings presented in Section 4, here we discuss
additional benefits and risks of chatbot-driven conversational
surveys.

\subsubsection{Quality Responses without Positivity Bias}

From Section 4, our study results clearly indicate that conversational
surveys draw out more relevant and richer user responses as well as
foster better user engagement. This is one of the obvious benefits of
using a chatbot for conducting surveys. However, existing research
shows that in a human-computer conversation, users might give more
positive responses because of their affections for humanized machines
\cite{thomaz2008teachable}. Any biased survey results, however, would
prevent researchers from discovering the truth and hinder decision
making. We thus examined whether our chatbot surveys caused any
potential positivity biases in key responses that would impact
business decisions.

In our study, our collaborator cared very much about the ``end
results''---participants' answers to a Likert scale question for
each game trailer on a scale of 1-5 \emph{``How interested are you in
  purchasing the game you just saw in this trailer''}. We compared the
ratings in the chatbot surveys (Trailer 1: M = 3.76, SD = 1.62;
Trailer 2: M = 3.75, SD = 1.63) and those in the Qualtrics surveys
(Trailer 1: M = 3.73, SD = 1.47; Trailer 2: M = 3.75, SD =
1.49). Controlling for demographics, game-playing time, and engagement
duration, an ANCOVA analysis showed that the rating differences were
not significant: Trailer 1: F(576) = 3.62, p = 0.06, $\eta_{p}^{2}$ = 0.01;
Trailer 2: F(576) = 2.18, p = 0.14, $\eta_{p}^{2}$ < 0.01. In other
words, the use of a chatbot did not influence the participants
to provide more positive ratings. 

Although the participants' ratings for the trailers did not seem to be
influenced by their affections for the chatbot, it is unclear whether
their overwhelmingly positive comments (Section 4) about their
perception of or attitude toward the chatbot was biased.

Nonetheless, our study reveals a key benefit of a chatbot survey:
eliciting richer and deeper participant responses while not causing
unwanted positivity biases.

\subsubsection{Coping with Survey Fatigue}

On average the participants spent over 20 minutes with the Juji
chatbot, which is considered extraordinarily long in an online survey
context \footnote{https://www.surveymonkey.com/curiosity/5-best-ways-to-get-survey-data/}. Since our results show that the
participants were willingly engaged with the chatbot longer and still
very positive about their experience, another benefit of a chatbot
survey seems to be in combating survey fatigue. To verify this
benefit, we examined the participants' response quality overtime,
since survey taking fatigue would negatively affect response
quality. We did not observe any quality degradation
  over time with a survey that lasted for about 20 minutes. On the
opposite, we found a question \emph{``Why do you give this score?''}
appeared very late in the survey even elicited better (i.e., more
relevant, specific, and clear) responses than most of the
\textit{what} questions asked in the middle of the survey. This might
be attributed to the question prompts used. Previous work shows that
\textit {why} questions encourage people to think deeper and offer
quality responses \cite{paul2006thinker}.

Moreover, the interactive nature of the Juji chatbot appeared to help
overcome survey-taking fatigue. Based on their comments, 42.2\% (119
out of 282) of participants explicitly mentioned that they really
enjoyed their chat with Juji and thought the experience ``cool'',
``entertaining'', and ``amazing''. Despite the potential novelty
effect as discussed below, the participants explicitly mentioned that
they liked their interaction with the Juji chatbot, which made the
survey not boring like typical online surveys. Because of the
interaction, certain participants even felt that they were talking to
Juji the chatbot as if they were chatting with a ``friend'', ``nice
guy'', or a ``brother''. It seems that such bonding encouraged the
participants to stay engaged and alleviated the survey-taking fatigue
often experienced in a traditional, static survey.

\subsubsection{Texting with a Chatbot on Mobile Devices}

With the widespread use of mobile devices, more and more people take surveys on their
mobile phones. Statistics provided by Survey Monkey showed that about
15\% of people nowadays take surveys on their mobile devices \footnote{https://www.surveymonkey.com/curiosity/are-people-completing-surveys-on-mobile-devices/}. However, studies showed when using mobile
devices, people's attention is limited and people are not willing to
type long sentences which is critical to elicit high-quality responses
especially to open-ended questions \cite{mavletova2013data,wells2014comparison,antoun2017effects}. On
the other hand, a Gallup
poll~\footnote{http://news.gallup.com/poll/179288/new-era-communication-americans.aspx}
shows that Americans under 50 use texting on their mobile phones as
the dominant communication method. Thus, another benefit of using a
chatbot survey seems to provide survey participants with a natural
communication form that they are most familiar with.

On the other hand, previous studies suggested the use of mobile
devices may prevent people from entering quality answers to open-ended
questions \cite{mavletova2013data,wells2014comparison,antoun2017effects}. We
thus investigated the relationships between the use of mobile devices
and participants' behavior in our study. Since our collaborator did
not collect participant's device information in the Qualtrics surveys,
we had only the device information for the chatbot survey
participants. Out of 282 chatbot survey takers, 95 (33.7\%) of them
used mobile devices, including cell phones and tablets. We analyzed
the relationships between participant's device use (mobile
vs. non-mobile) and various response quality metrics and participant
engagement metrics (Section ~\ref{sec:measures}). For each metric, we
constructed an ANCOVA analysis controlling for demographics,
game-playing time, and engagement duration. We found no significant
associations between the usage of mobile devices and the participant's
response quality nor engagement. This is very encouraging, since our
study results show no evidence that the use of mobile devices would
adversely influence participants' behavior in a chatbot survey like ours. 

\subsubsection {Understanding User Characteristics}

One of the purposes of conducting surveys is to understand the
characteristics of target participants \cite{muller2014survey}. Understanding participant characteristics
has multiple benefits. For example, they can be used to help explain
survey results. In addition, they can be used to effectively guide a
conversation (e.g., persuading a user based on his/her personal
characteristics \cite{hirsh2012, petty1981}). Unlike an ordinary
chatbot, another unique feature of the Juji chatbot is its ability
to analyze a user's text input on the fly and infer the user's
characteristics \cite{zhoutiis2019}. In the current study, we
explored this feature of the Juji chatbot preliminarily. Near the end
of each chatbot survey, the chatbot analyzed a participant's text
responses given in the survey and automatically inferred his/her key
gamer characteristics.

For example, one participant was told that she had the characteristics
of a \emph{``social gamer''}, who is extroverted, friendly and enjoys
playing games with friends. In contrast, another participant was
informed that he possessed the characteristics of a \emph{``complete
  gamer''}, who is very driven to achieve all game milestones. To
verify the accuracy of its inference, the chatbot also asked each
participate to rate the accuracy of their inferred gamer
characteristics on a scale of 1-5, 1 being completely off and 5 being
very accurate. The results showed that M = 4.55, SD = 0.65, which was
very encouraging.

To better explain the participants' ratings about their interest in
purchasing a game, our collaborator wanted to know how participants'
inferred gamer characteristics were related to their ``interest to
purchase'' rating of the two respective game trailers. To do so, we
performed a regression analysis on the inferred participants' gamer
characteristics and their game purchase interest. Our results showed
that ``social gamers'' preferred game 1---a shooting strategic game
with many action elements, while `` passionate gamers'' favored the
second game---an action-oriented strategic game with a compelling
storyline. 

Although in this study our investigation of the relationships between
participants' inferred characteristics and their purchase interest is
very preliminary, our study suggests another benefit of chatbot-driven
surveys. Specifically, a chatbot can potentially achieve a ``two birds
with one stone'' outcome: eliciting information from survey
participants and using the elicited information to infer participants'
characteristics at the same time. This would reduce survey-taking time
since separate surveys (e.g., gamer type survey) intended to
understand participants' characteristics may no longer be
needed. Moreover, the inferred participant characteristics would help
gain deeper insights into the collected information.

\vspace{-2mm}
\subsubsection {Chatbot Addiction}
Numerous studies show that people may become addicted to powerful and
omnipresent technologies. For example, research shows that the
omnipresence of mobile phones has caused mobile phone use addiction
~\cite{park2005}. Users' own characteristics, such as personality
traits, could even predict addiction behavior
~\cite{bianchi2005}. Likewise, as chatbots' capabilities become more
advanced and their uses become more ubiquitous, the bonding between
humans and machines may grow stronger and potentially lead to certain
unwanted effects, such as chatbot addiction. Compared to other
technology addictions, chatbot addictions would be more likely to
occur and harder to overcome due to the anthropomorphic and personal
nature of human-chatbot interactions. It is thus important for
chatbot designers and developers to be aware of potential usage
behaviors including addiction behavior, and consider proper behavior
detection and prevention mechanisms as part of the chatbot design. For
example, research shows that certain user characteristics, such as
extroversion and self-esteem, could predict problem uses of technology
including technology addiction ~\cite{bianchi2005}. Chatbot designers
and developers may borrow such findings to detect potential addiction
behavior and deter users from overengaging with a chatbot.

\vspace{-2mm}
\subsubsection{User Privacy and Control}
Our study results reported in the last section along with other
  studies suggest that users are willing to disclose sensitive,
  personal information to a chatbot (e.g., ~\cite{sundar2019machine}). In
  addition, a recent study found that a chatbot could gain a user's
  trust if it informs users that their data will be securely stored
  \cite{folstad2017chatbots}. While gaining user trust helps a chatbot
  elicit authentic information in a survey context, these findings
  reveal potential risks of the malicious uses of chatbots. Such
  chatbots can manipulate users to gain their trust and steal their
  sensitive, personal information. Moreover, even benevolent chatbots
  might elicit unnecessary sensitive information since users tend to
  disclose more to a chatbot than to a human \cite{lucas2014s}. Those
  unnecessary self-disclosures may expose users to privacy leakage or
  identity theft risks if data breaches occur.

  Therefore, proper chatbot design and evaluation guidelines should be
  in place to allow a chatbot to accomplish its task while protecting
  user privacy. For example, in our study certain participants
  disclosed where they live and what kind of job they do when they
  were asked to introduce themselves. In such a case, the chatbot
  could be designed to warn the participants not to disclose
  personally identifiable information. In addition, the chatbot could
  show a participant what information has been gathered and provide
  the participate the options to control the use of the information
  (e.g., obfuscating parts of the information as needed
  ~\cite{wang2015veilme}). Not only do these privacy controls help guard user
  privacy, but they also improve user engagement and satisfaction
  \cite{vaccaro2018illusion}. However, such user controls may
  interfere with the collection of authentic information (e.g.,
  authentic patient information for determining proper treatment) as
  certain parts of data might be obfuscated or removed.  Therefore,
  future studies are needed to investigate how to reach a balanced
  design of chatbot-driven surveys that can protect user privacy while
  guard information validity.

\subsection{Study Limitations}

Our current study has several limitations, including flaws in the study
operations and scope of the results.

\subsubsection {Study Controls}
\label{sec:studycontrols}

Just like any field studies, our study was limited by practical
constraints imposed by real-world operations. One constraint was that
we had to use a panel company to find the qualified participants (over
18 years old, hard-core gamers) for the purpose of the study. We had
no control over how the two target audience groups were selected, how
recruiting messages were sent out, or how the reward was
determined. For example, the initial recruiting messages sent to the
two target groups were the same except one containing a link to
Qualtrics and the other to the chatbot. During the first couple of
days, such a message however caused a 30\% abandon rate in the chatbot
survey. Our log data indicated that a large number of participants
abandoned the survey as soon as the chat screen appeared. Our guess
was that they thought this was a regular survey per the recruiting
message but the chat screen did not look like a typical online survey
they were used to, which made them abandon the survey immediately. We
thus asked the panel company to revise the recruiting message, which
explicitly informed the participants that they would chat with an AI-based 
chatbot in this survey. The later revised message may adversely impact the 
response rate. The participants may choose not to take the survey because 
of their familiarity with the novel chatbot-driven survey. Although the revised 
message reduced the abandon rate dramatically, the overall completion rate given
by the panel company (Equation~\ref{fun:completion}) was affected by this incident. 

\subsubsection {Study Audience and Scope}

Since our study aimed at understanding gamers' thoughts and feelings
about newly released game trailers, this gamer-focused study might
limit the applications of our results to other populations. Although
research shows that the newer generation grows up playing video games,
gamers often form their own beliefs and perhaps are more open and
receptive to new technologies such as chatbots
~\cite{Carstens2004}. As mentioned in Section 4, our analyses
showed that game-playing time indeed contributed to the difference in
response relevance. It is thus unclear whether our results would hold
for non-gamer populations. Additionally, our study is on participants'
opinions of game trailers of popular games, it is unclear whether our
results would hold for other types of surveys, for example, employee
engagement surveys or market research surveys for more ``mundane''
products (e.g., household products) or services (e.g., banking
service). Although none of the analyses showed that participants'
\emph{age} contributed to the differences in two survey methods,
two-thirds (2/3) of the participants were between the age of 18-34
(Fig ~\ref{fig:age}). Therefore, it is also unclear whether our
results would hold for populations in other age groups.

  Additionally, our study focused on investigating the
  use of chatbots for collecting user free-text responses to
  open-ended questions. As mentioned in the Introduction, open-ended
  questions are an important way to elicit important user insights and
  are widely used in web-surveys \cite{reja2003open}. However,
  eliciting quality responses to open-ended questions is very
  challenging in a typical online survey since participants are often
  not motivated to provide detailed, rich input. Moreover, no existing
  survey platforms provide tools to facilitate the collection of user
  responses to open-ended questions. Therefore, our study focused on
  examining the use of an AI-powered chatbot and its effect on surveys
  with open-ended questions, hoping to find new ways to aid in such
  surveys. However, it is unknown how the use of an AI-powered chatbot
  would aid in surveys with choice-based questions. Although one
  recent study shows that a chatbot survey exhibited less satisficing
  behavior on choice-based questions \cite{kimchi2019}, it is unclear
  whether such a chatbot handles any user digressions (e.g., a user
  does not give an answer by selecting a choice) as in our
  study. Further studies are definitely needed to examine the
  effectiveness of chatbots in eliciting information through all types
  of questions. 

\subsubsection{Novelty Effect}

In recent years, chatbots or intelligent agents have been widely
adopted in our daily lives \cite{dunnwe,cui2017superagent}. In a
business context, chatbots have also been used in a wide range of
applications, from job interviewing ~\cite{li2017confiding} to serving
as a workplace companion ~\cite{williams2018supporting}. Nonetheless,
it is still uncommon to use chatbots for conducting lengthy,
conversational surveys that mainly consisted of open-ended questions
as we did in our study. Furthermore, few chatbots used in a survey
context ~\cite{tallyn2018ethnobot, kimchi2019} have any conversation
skills capable of handling diverse and complex user interactions as
the Juji chatbot did.

As the first study of its kind, our study setting was a novelty to the
majority of the participants. In particular, two novelty factors
presented in our study might have affected participant behavior and
biased our study results: (1) the form of the conversational survey
itself with mainly open-ended questions; and (2) the rich conversation
skills of the Juji chatbot. Since we could not control for the novelty
effect in our current study design, we do not know the contributions
of the novelty factors. This is certainly one of the limitations of the study.

While we are planning longitudinal studies to examine the influence of
the novelty effect, here we briefly discuss the potential effect of
the two novelty factors mentioned above. Just as any novel technology,
the novelty effect may wear off as chatbots become a norm. In our
case, as machine-driven conversational surveys become more common, the
effect caused by the first novelty factor is most likely to wear off,
similar to the fact that online surveys are now a norm in lieu of
pencil and paper surveys. However, as chatbots' conversation
capabilities become more powerful, the second novelty factor would
continue influencing user behavior beyond their novelty.  This is
because chatbots' inherent interactive features, such as
acknowledging, prompting, and probing, directly influence
human-machine interactions ~\cite{bohus2009ravenclaw}.

In our view, the influence of chatbot technologies on user behavior
would be similar to that of mobile technologies. Initially, user
adoption of mobile devices might be attributed to the novelty of
the technology. However, as the mobile technologies advance and can offer
users a wide range of functions, their influence on user behavior
is beyond their novelty. Analogously, as AI agents advance their
capabilities, e.g., their better understanding of user responses and
initiating more adaptive questioning (see ``Going beyond surveys''),
chatbots will greatly influence user behavior beyond their
novelty. Just like mobile technologies, they may even induce unwanted
effects (e.g., chatbot addiction, as discussed above).

\subsubsection {Effect of Individual Chatbot Skills}

Our study was specifically designed to investigate the use of a chatbot \emph{as a
  whole} for conducting surveys and compare its outcomes (i.e.,
resulted survey quality and user engagement) with that of a
traditional online survey. From analyzing the chat transcripts, we
observed that the Juji chatbot used a rich set of conversation skills
to interact with a participant in a conversational survey, such as
information prompting, answer probing, and handling side-talking
(Table ~\ref{tab:conversationalfeatures}).

Below is a chat transcript showing two of the features, prompting and probing.

\begin{table}[H]
\centering
\begin{tabular}{p{10mm}p{110mm}}
\hline
Juji: & What is your favorite type of games?                  \\
\rowcolor[HTML]{EFEFEF} 
User: & sports and adventure strategies                       \\
Juji: & Got it, I'm interested in hearing a bit               \\
      & more. Please?  \textless{}Prompting\textgreater{}     \\
\rowcolor[HTML]{EFEFEF} 
User: & I love video games since I was a very                 \\
\rowcolor[HTML]{EFEFEF} 
      & fanatical child and I am currently                    \\
\rowcolor[HTML]{EFEFEF} 
      & following the latest trends                           \\
Juji: & Ok. Thanks.                                           \\
Juji: & Why do you like it?  \textless{}Probing\textgreater{} \\
\rowcolor[HTML]{EFEFEF} 
User: & it's my favorite pastime and latest technology        \\ \hline
\end{tabular}
\end{table}

In fact, from the 282 completed chat transcripts, on average, each
participant was prompted by the chatbot 2.9 times for more
information. The participants complied with the prompting request and
offered more information 71\% of the time. In addition to information
prompting, the chatbot asked follow-up probing questions as shown
above. By the transcripts, the chatbot asked at least one follow-up
question to 173 out of 282 (61\%) participants who completed a
conversation survey. The participants answered a follow-up question
89\% of the time.

Moreover, the chatbot engaged the participants in side-conversations in 32 out of 282 (11\%)
conversations, such as answering the participants' questions or
dealing with their unwillingness to provide answers. Below is a chat
excerpt, which shows how the chatbot handled the side conversations and was
persistent in getting an answer from the participant.

\begin{table}[h]
\centering
\begin{tabular}{p{10mm}p{110mm}}
\hline
Juji: & What do you not like about it?                \\
\rowcolor[HTML]{EFEFEF} 
User: & Why are you repeating the question            \\
Juji: & Why not? :) I am pretty sure this is for your own good since I am a responsible A.I. \textless{}Side Talking\textgreater{}\\
Juji: & Well, let's just get going.                   \\
Juji: & What do you not like about it?                \\
\rowcolor[HTML]{EFEFEF} 
User: & Woah you understood that smart ai             \\ \hline
\end{tabular}
\end{table}

The participants' behavior observed in our study suggests that
conducting a conversational survey is more than merely asking survey
questions one by one especially when open-ended survey questions are
involved. As shown in the chat transcripts, the chatbot's various
conversation skills, such as the abilities to prompt for more
information and probe answers whenever appropriate, contributed to the
survey success. In addition, the chatbot's abilities to handle diverse
conversation situations, such as user excuses, questions, and bogus
responses, also contributed to the survey success. Intuitively, each
of the chatbot conversation skill might have contributed to the survey
success differently.

However, the data collected in this study is inadequate to tease apart
and quantify the contribution of prompting or any other chatbot
interaction feature due to the highly adaptive chatbot behavior, which
was intended to offer the best possible user experience in this
study. For example, in the study, the triggering of prompting depended
on the nature of user input and the context (e.g., the type of
question asked). For example, a participant who gave an irrelevant
response such as gibberish input was prompted to provide a relevant
answer, while a participant who gave a relevant but terse response was
prompted to give more detailed, specific input. The question context
also determined the triggering of chatbot's interaction features. For
example, in the question \emph{``what's your reaction to the trailer
  you just watched''}, the chatbot would prompt a participant for more
information if the participant's response is very short. In contrast,
when asked \emph{``what's your favorite game''}, a participant's short
response might be sufficient and would not trigger the chatbot
prompting.

In a conversation, different open-ended questions require different
relevant answers. Additionally, to balance the amount of information
elicited and the time required to complete a survey, the chatbot could
not afford to prompt a participant for more information in every
open-ended question. Moreover, the multiple interaction features of
the chatbot often worked in combinations to nudge users to provide
quality input. As shown in the chat transcript below, one user's
input triggered a prompting, a probing, and another prompting. Since a chatbot's
interaction features have both benefits and risks,
rigorously quantifying the contribution of each chatbot conversation
skill would be valuable. However, this would require additional, fully
controlled experiments beyond the scope of the current study.

\begin{table}[h]
\centering
\begin{tabular}{p{10mm}p{110mm}}
\hline
Juji: & What is your favorite type of games (e.g. first person shooter, multi-player, open world, strategy, sports/racing, battle royale, etc.)? Please give an example of such games.                \\
\rowcolor[HTML]{EFEFEF} 
User: & sports are my favorite            \\
Juji: & Oh,  go on... \textless{}Prompting\textgreater{}\\
\rowcolor[HTML]{EFEFEF} 
User: & i like fighting and racing too          \\
Juji: & Thanks. Why do you like it? \textless{}Probing\textgreater{}\\
\rowcolor[HTML]{EFEFEF} 
User: & its fun to play          \\
Juji: & Well,  anything else? \textless{}Prompting\textgreater{}\\
\rowcolor[HTML]{EFEFEF} 
User: & mario kart is one of my favorites          \\\hline
\end{tabular}
\end{table}

\subsection{Design Implications}

Here we present several design considerations for creating effective
chatbots for conducting conversational surveys.

\subsubsection {Active Listening}

Quantitatively, our results showed that the Juji chatbot elicited
significantly higher quality responses and significantly more
participant engagement. Qualitatively, over half of the participants
(67.4\%) expressed their highly positive experience with the
chatbot. Our qualitative analysis of the chat transcripts also
revealed that the chatbot's rich conversation skills play a critical
role in the success of a conversational survey. These skills enable
the chatbot to \emph{actively listen} to its users and make the users
feel heard, which not only delighted the users, but also nudged the
users to contribute more quality answers. Our findings are consistent
with other study findings that active listening improves communication
effectiveness in text-based communication \cite {Bauer2010} as well
as in information elicitation \cite{Louw2010}. In other words, 
adopting a chatbot with active listening skills helps achieve survey
effectiveness.

\subsubsection{Intervening Early}

In our initial analysis, we noticed that the quality of a
participant's response to the first survey question (self-intro) seems
a barometer of the quality of his/her overall responses. For example,
the participants who gave a gibberish response to the first question
also gave gibberish responses to all other questions. We thus analyzed
the correlation between the quality of the first responses to that of
overall responses. We found a \textit{significant} correlation between
the \textit{relevance} scores of participants' first responses and
that of their overall responses: r = 0.78, N = 582,
p<0.01**. Similarly, a significant correlation exists between the
response quality index of the first responses and that of overall
responses: r=0.70, N=582, p<0.01**.

These results imply that if a participant diligently answers the first
question, it is highly likely that s/he would do so for the entire
survey. In this light, the first open-ended question could be used as
an effective screener to assess a participant's willingness to do a
survey. Since prior work shows that machine "intervention" could be
effective (e.g., preventing fake answers in an interview
\cite{law2016fake}), a chatbot could even intervene if it detects a
participant's unwillingness to answer the first question. Such early
intervention benefits information collectors as well as participants
since the participants can be reminded of their responsibilities and
expectations.

\subsubsection{Mixing Chatbot with Qualtrics} 

Our collaborator, the market research firm, was satisfied with the
study results and wishes to use a Juji-like chatbot in their future
studies especially for eliciting qualitative responses. They, however,
want to explore a hybrid use of a chatbot with a typical online survey
where they can use a platform like Qualtrics to ask complex
quantitative questions (e.g., Matrix rating) while leveraging a
Juji-like chatbot for open-ended questions. We see two ways of
integrating the two methods: embedding the chatbot into a Qualtrics
survey or vice versa. On the one hand, it might be more effective to
start with a chatbot and then move to a Qualtrics survey because of the
chatbot's interactive features including its potential ability to
intervene. On the other hand, it may be better to start with
quantitative questions on Qualtrics and then transition to the
chatbot, which would help battle survey taking fatigue. It would be
interesting to study different hybrid models and their effect on
survey results and participant experience.

\subsubsection{Creating Empathetic and Responsible AI Agents Beyond Surveys}

From this study, we observed a chatbot's abilities to conduct surveys
or structured interviews based on a set of pre-defined questions and
the order of the questions. The conversation capabilities of the
chatbot demonstrated in our study suggest the natural next step to
advance its capabilities for wider applications. In particular, we see
the opportunity to use a chatbot for replacing structured phone
interviews. It is also possible to develop a chatbot for
semi-structured interviewing. Starting with an interview guide, a
chatbot will ask open-ended questions, interpret user answers, and
\textit{automatically} come up with follow-up questions to drill down
on interesting ideas that emerged in the conversation and uncover hidden
insights. To achieve this goal, several key advances still need to be
made, including automatically formulating critical questions based on
a participant's response, as suggested by the Socratic questioning
method \cite{paul2006thinker}.

Moreover, our results including participants'
comments also indicate the possibility of creating a new generation
of AI agents, which can deeply understand users (e.g., one user is
open-minded and easy going while another is analytical and cautious)
and proactively guide users based on the conversation and the unique
characteristics of the users (e.g., customizing survey questions for
analytical participants). Not only can such AI agents exhibit empathy
during their interactions with users and gain user trust, but they
can also be made responsible for their actions (e.g., guiding users
to provide authentic information while guarding user privacy).

In our study, although the average engagement duration was only 20
minutes, certain participants already felt a personal connection
with the Juji chatbot. We envision that empathetic and responsible
agents can better bond with users and accomplish a wide range of
tasks beyond conducting surveys. For example, they can serve as
personal well-being coaches, career counselors, and personal
caretakers. Such agents will also push the boundary of relational
agents \cite{bickmore2011relational} and help achieve the goal of creating true
human-machine symbiosis \cite{licklider1960man}. It would be exciting to
investigate the uses of and effect of such AI agents on people's
daily lives.

\section{Conclusions}
We reported a field study that compared the outcomes of a
chatbot-driven survey and that of a typical online survey. The study
involved about 600 participants, half of them took a chatbot survey on
Juji (\href{https://juji.io/}{juji.io}) and the other half filled out a form-based online
survey on Qualtrics (\href{https://www.qualtrics.com/}{qualtrics.com}). Compared to the form-based
Qualtrics survey, the AI-powered chatbot survey was a conversational
survey during which the chatbot provided interactive feedback to
free-text responses, prompted for information, probed answers, and
handled various social dialogues whenever appropriate. Through an
in-depth analysis of over 5200 free-text responses collected from the
study, our results showed that the participants who completed a
chatbot survey provided significantly more relevant, specific, and
clear free-text responses than their counterparts did in a Qualtrics
survey. They were also more willing to spend time with the chatbot,
provide longer responses, and disclose more information about
themselves. 190 (67.4\%) of the participants who engaged with the
chatbot also expressed their positive experience and willingness to
take surveys in a chat format.

Given our study results and the simplicity of creating and deploying a
chatbot survey like the one used in our study, our work suggests a new
and promising method for conducting effective surveys especially for
the purpose of collecting free-text responses to open-ended questions
and overcoming survey taking fatigue. With the increasing use of
chatbots, our results also present important design implications for
creating and employing chatbots for survey success. In particular,
chatbots should be equipped with active listening skills to guide
participants in a conversational survey, which will elicit higher
quality responses and deliver better engagement experience. Chatbots
can also be used to deliver early interventions at the beginning of a
survey, which would encourage quality responses and prevent
cheating. Moreover, it is desirable to mix the use of a chatbot and a
typical online survey to accommodate different survey goals and
maximize survey success.



\begin{acks}
This work is in part supported by the Air Force Office of Scientific Research under FA9550-15-C-0032. We would like to thank our collaborators, Alex Dillon and Michael Cai at Interpret, and all our participants in our study. We also want to thank our Associate Editor, Dr. Darren Gergle and all anonymous reviewers for their valuable feedback.
\end{acks}
\bibliographystyle{ACM-Reference-Format}
\bibliography{proceedings-sigchi}

\end{document}